\documentclass[a4paper,11pt]{article}
\usepackage{aaskaiid}
\usepackage{comment}
\usepackage{aas_macros}
\usepackage{orcidlink}
    \usepackage[utf8]{inputenc}
    \usepackage[T1]{fontenc} 
    \usepackage{newunicodechar}
    \newunicodechar{≫}{\ensuremath{\gg}}

\title{Coronal Magnetography using Spectropolarimetry with SKA Telescopes}
\ShortTitle{Coronal magnetography}

\author[1]{Deepan Patra \orcidlink{0009-0003-8133-6621}}
\ShortName{Patra et al.} 
\author[1]{Puja Majee \orcidlink{0000-0002-2711-2366}}
\author[1]{Soham Dey \orcidlink{0009-0006-3517-2031}}
\author[2,3, 4]{Devojyoti Kansabanik \orcidlink{0000-0001-8801-9635}}
\author[1]{Divya Oberoi \orcidlink{0000-0002-4768-9058}}
\author[5]{Anshu Kumari \orcidlink{0000-0001-5742-9033}}
\author[1]{Surajit Mondal \orcidlink{0000-0002-2325-5298}}
\author[6]{Ketaki Deshpande \orcidlink{0000-0001-6861-6328}}
\author[5]{Divya Paliwal \orcidlink{0009-0001-7689-0084}}

\affiliation[1]{National Centre for Radio Astrophysics, Tata Institute of Fundamental Research, S. P. Pune University Campus, Pune, India, 411007}
\affiliation[2]{Instituto de Astrofísica de Andalucía-CSIC (IAA-CSIC), Granada, Spain}
\affiliation[3]{NASA Jack Eddy Fellow, University Corporation for Atmospheric Research, 3090 Center Green Dr., Boulder, CO, USA, 80301}
\affiliation[4]{Johns Hopkins University Applied Physics Laboratory, 11001 Johns Hopkins Rd, Laurel, MD, USA 20723}
\emailAdd{deepanpatra1999@gmail.com}
\emailAdd{pujamajee@gmail.com}
\affiliation[5]{Udaipur Solar Observatory, Physical Research Laboratory, Dewali, Badi Road, Udaipur - 313001, Rajasthan, India}
\affiliation[6]{Center for mathematical Plasma Astrophysics (CmPA), KU Leuven, Celestijnenlaan 200B, 3001 Leuven, Belgium}

\abstract{ 
The solar coronal magnetic field drives nearly every aspect of solar phenomena and activity -- from flares, coronal mass ejections, and solar wind that governs space weather to the much weaker nanoflares. 
These magnetic fields are routinely measured at the visible surface of the Sun, the photosphere. 
However, detailed and direct measurements of the magnetic fields in the solar atmosphere, particularly in the coronal layer, have remained rather limited. 
Mostly, these are estimated from vector magnetic field measurements at photospheric heights through different extrapolation models. 
In the case of the corona, these extrapolations lack observational constraints from the corona, especially during periods of intense activity when magnetic structures evolve rapidly.
Measurements of coronal magnetic fields from observations, therefore, remain one of the most crucial and unresolved challenges in solar and space-weather research. Radio observations of the Sun hold considerable potential in this regard. 
Observations of diverse emission mechanisms, ranging from plasma emissions at lower frequencies to thermal Bremsstrahlung and gyro-resonance at higher frequencies, provide multiple avenues to probe the coronal magnetic fields, unique at radio wavelengths.
SKAO, with its broad frequency coverage (0.05 to 15 GHz), will allow us to probe  
wide range of coronal layers through unprecedented high-fidelity polarimetric imaging at high temporal, spectral, and spatial resolutions.
This chapter details how the coronal magnetic field measurements can be achieved through spectro-polarimetric imaging of the Sun with the SKAO.
}

\begin{document}
\maketitle

\section{Introduction}
The magnetic field in the solar atmosphere plays a central role in shaping plasma structures and dynamics, as well as in the manifestation of solar activities. In contrast to the photosphere and the chromosphere, the magnetic pressure in the corona dominates the plasma pressure. As a result, the magnetic field determines the distribution and plasma structures of the corona.
Moreover, the magnetic energy accumulation and release drive energetic phenomena, such as flares, acceleration of solar energetic particles (SEPs), coronal mass ejections (CMEs), which have a significant impact on the heliosphere and the space weather (SpWx) around Earth. 
Hence, understanding and quantifying the coronal magnetic fields and their temporal evolution is important for studying the energetics in the corona, including the storage and release of magnetic energy. 

Until recently, in-situ measurements of the coronal magnetic field were not possible. With the closest approach of the Parker Solar Probe \citep[PSP;][]{Raoutfi2023} to $\sim$9.8$R_\odot$, it is now possible to obtain in-situ magnetic field measurements of the middle and outer corona. However, these measurements only provide localized information. Hence, the global coronal magnetic fields measurement entirely relies on remote sensing observations -- inferring magnetic fields from the influence of it on the electromagnetic radiation from the corona. Magnetic field measurement via the Zeeman effect on the spectral line emission from the ion is routinely used for the solar photosphere \citep{Lagg2017}. However, it becomes ineffective for coronal magnetic field measurements because the coronal magnetic fields produce extremely small Zeeman splittings, far below the line broadening caused by thermal motions of the coronal plasma. Moreover, the corona is optically thin, so emission integrated along the line-of-sight (LoS) leads to the signal being mixed and cancelled. 

There are several techniques for coronal magnetic field measurement across the electromagnetic spectrum (as listed in Table 1 of \citet{Gibson2016}). Radio wavebands provide unique stands, particularly in providing both quiet time and active time magnetic field measurements over a wide range of heights.
Solar radio emissions can be divided into two parts --  the highly variable, bright `active' emission, and comparatively faint, slowly varying `quiet' emission. 
To probe the ever-evolving solar corona, one requires wideband observations with very high spectro-temporal resolution and a high dynamic range to capture both faint and bright features simultaneously. 

In this chapter, we first describe the challenges in measuring the coronal magnetic field and the role of radio observations in overcoming those, in Section \ref{sec:challenge}. This is followed by a brief discussion of the different remote-sensing techniques in the radio band to probe the coronal magnetic field in Section \ref{sec:mag_field_measurement}. Section \ref{sec:role_of_ska} discusses how the SKAO will play a crucial role in coronal magnetography and is followed by a Summary in Section \ref{sec:summary}.

\section{Overcoming Challenges in Coronal Magnetography using Radio Observations}\label{sec:challenge}
The coronal magnetic fields are intrinsically three-dimensional (3D), which makes their measurement a significantly different problem compared to two-dimensional (2D) photospheric magnetic fields. At the photosphere, magnetic field strength can vary from several Gauss to several thousand Gauss. Hence, the Zeeman effects in the optical spectral lines are measurable and provide measurements of the photospheric surface magnetic field. In the solar corona, magnetic fields weaken rapidly with height, while the low-density plasma produces optically thin spectral lines. Consequently, Zeeman polarization signals originating from regions with different magnetic field orientations cancel along the line of sight, yielding a negligible net polarization. Coronal Zeeman splittings are far smaller than thermal Doppler widths of coronal emission lines due to the higher temperature than the photosphere. 
The signal-to-noise ratio (SNR) is also limited due to the faintness of coronal emission lines, making precise spectro-polarimetric detections extremely challenging, even in infrared lines where the Zeeman sensitivity is enhanced \citep{lin2004coronal}. Recent advances, such as high-sensitivity infrared coronal polarimetry with instruments such as DKIST \citep{rimmele2020daniel} have enabled localized detections of Zeeman signatures in coronal Fe XII lines, marking a promising step toward direct magnetic mapping in the low corona \citep{schad2024mapping}. However, these measurements remain restricted to bright, low-altitude regions under highly favourable conditions. 

\begin{figure}[!h]
    \centering
    \includegraphics[width=0.95\linewidth]{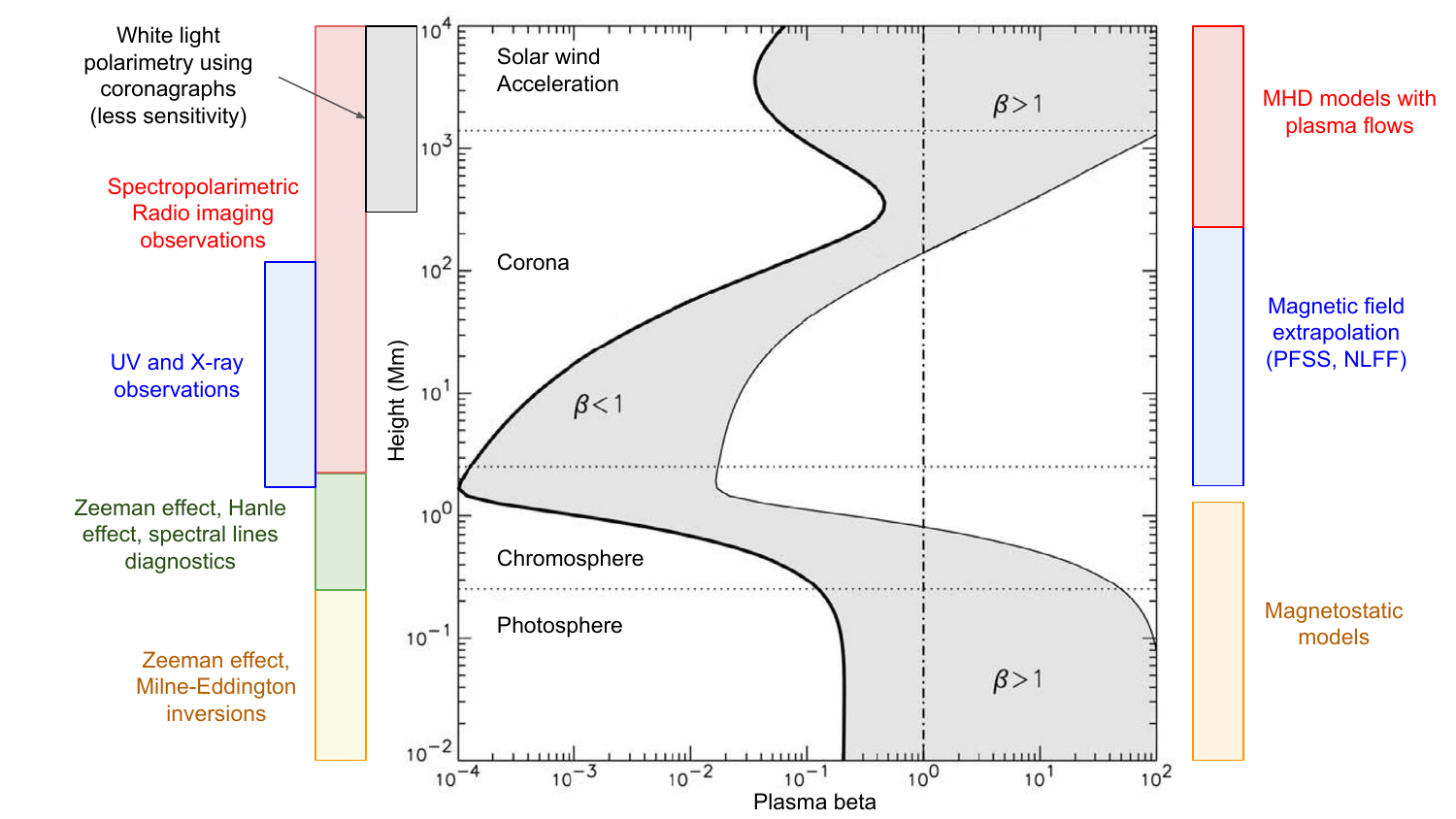}
    \caption{This schematic shows the variation of plasma $\beta$ parameter with height above an active region. The right side shows the current numerical models used for magnetic field modeling at different heights. On the left side are the observations from which magnetic field information can be obtained at different heights. The original plot was taken from \citet{gary2001plasma}.}
    \label{fig:plasma_beta}
\end{figure}

Numerical modelling offers a way to understand the structure and strength of the coronal magnetic fields. These models typically rely on extrapolations of the photospheric magnetic field to reconstruct the three-dimensional coronal field under various physical assumptions \citep{wiegelmann2021solar}. The simplest of them is the potential field source surface (PFSS) model, which assumes a current-free, static corona.
PFSS model has been widely employed to study the global magnetic topology and open field regions such as coronal holes \citep{schatten1969model}. It imposes a spherical potential source surface, typically at $\sim$2.5$R_\odot$, where all the field lines are constrained to be radial. This assumption makes the model incapable of capturing the complex magnetic field over active regions.
More advanced linear and nonlinear force-free field (LFFF and NLFFF) models, which include electric currents, provide improved representations of active regions, capturing the non-potential structures associated with flares and filaments \citep{wiegelmann2008nonlinear}. Beyond these, data-driven and data-constrained magneto-hydrodynamic (MHD) simulations have emerged as powerful tools for reproducing the temporal evolution of coronal magnetic fields and plasma dynamics  \citep{ van2014alfven, Perri22, Perri23, daei2023modeling,downs2025near}.
Despite their widespread use, coronal magnetic field models suffer from several fundamental limitations that constrain their accuracy and physical realism. Most extrapolation techniques, such as the PFSS and NLFFF models, assume static or quasi-static, force-free conditions while using photospheric magnetic field measurements as lower boundary conditions. The most important issue in these approaches is that the photospheric high-$\beta$ \footnote{$\beta=\frac{\text{Plasma Pressure}}{\text{Magnetic Pressure}}=16\pi\xi nK_BT/B^2$, where $\xi=1$ for corona and $1/2$ for photosphere, n is particle density, T is temperature and B is magnetic field.} plasma conditions do not match the low-$\beta$ coronal conditions, as evident from Figure \ref{fig:plasma_beta}, making it physically inconsistent to directly use the photospheric field as a boundary condition for coronal models \citep{wiegelmann2017coronal}. Moreover, the photospheric plasma dominates the dynamics, and hence, the force-free assumption on the photosphere is not valid. While MHD models provide a global, self-consistent coronal magnetic field, they are limited by the resolution and cadence of photospheric magnetograms. Hence, such models cannot capture small-scale, rapidly evolving features such as reconnection sites or flare-driven dynamics \citep{Brchnelova23}. 

Radio observations can be used to fill the gap between measuring and modeling the coronal magnetic field \citep[see review by][]{alissandrakis2021radio}. Radio emission processes such as gyroresonance and gyrosynchrotron radiation depend directly on the local magnetic field strength and orientation, enabling spatially resolved measurements of coronal magnetic fields in active regions, where optical or EUV techniques fail \citep{gary1994coronal}.
Moreover, the propagation effects (e.g., quasi-transverse (QT) propagation) affect the observed polarization properties of radio waves. This encodes additional information about the magnetic field’s direction and plasma parameters along the line of sight. Additionally, plasma emissions, such as solar radio bursts, serve as valuable probes of coronal magnetic field strength and topology, as their generation, propagation, and polarization characteristics are intrinsically linked to the local magnetic field configuration. Hence, it can provide an indirect estimate of the magnetic field strength at low frequencies (<500 MHz) i.e., higher heights in the corona.
By comparing synthetic radio maps derived from PFSS, NLFFF, or MHD models with spectro-polarimetric radio images, quantitative constraints can be placed on the model parameters. Figure \ref{fig:plasma_beta} schematically summarizes the observational probes and numerical models used to measure and model magnetic fields across the solar atmosphere as a function of height.
\section{Magnetic field measurement using Radio observations}
\label{sec:mag_field_measurement}

\subsection{Free-free emission}
\subsubsection{Emission mechanism}
Under the electric fields of ions, electrons produce continuum thermal free-free (bremsstrahlung) emission. At radio wavelengths, this mechanism dominates the quiet Sun and also many active region conditions, with the plasma remaining optically thin except in dense loops or low-lying structures. The observed brightness temperature ($T_B$), therefore, becomes a reliable tracer of the thermal plasma distribution, provided the temperature profile is constrained independently.
The free–free absorption coefficient arises from bremsstrahlung theory \citep{kramers1923xciii}, while \citet{dulk_ff} provided convenient approximations tailored for solar radio applications. Recently, an extended formulation that correctly accounts for the varying Gaunt factor and different ionisation states has been presented by \citet{fleishman_ff}. Assuming a Maxwellian distribution, no magnetic field and unit refractive index, one can write the absorption coefficient as:
\begin{equation}
    \kappa_{ff}= \frac{8e^6}{3\sqrt{2\pi}ck_Bm_e^{3/2}}\frac{n_e^2\ln{\Lambda_c}}{\nu^2T^{3/2}}(1+\zeta(T,\nu))
\end{equation}
where the constants have their usual meaning. The ionization term, $\zeta(T,\nu)$ is given by
\begin{equation}
    \zeta(T,\nu)=\sum_{i=2}^NZ_i(g_iZ_i-1)\frac{n_i}{n_e}
\end{equation}
here, $g_i=G_i/G_1$ is the ratio of Gaunt factor of i-th ionized ion ($G_i$) to the singly ionized ion ($G_1$).
Furthermore, the Coulomb logarithm with the correct boundary temperature \footnote{ The correct transition temperature of 0.89125 MK was derived by \citet{fleishman_ff}. Using the previous value of 0.2 MK as derived by \citet{dulk_ff} introduces a small ($\sim$3\%) error.} can be written as:
\begin{equation}
    \ln\Lambda_{C} =
\begin{cases}
17.718414 + \ln\!\left(\dfrac{T^{3/2}}{Z_i}\right) - \ln \nu, & T < 0.89125\text{ MK} ,\\[6pt]
24.569056 + \ln T - \ln \nu, & T > 0.89125\text{ MK} .
\end{cases}
\end{equation}

Although free–free emission is intrinsically unpolarized, it acquires weak circular polarization in a magnetized plasma. This arises due to the difference in absorption of extraordinary (x) and ordinary (o) modes in anisotropic coronal plasma in the presence of magnetic field. Using the treatment prescribed in \citet{zlotnik_ff}, one can write:
\begin{equation}
    \kappa_{ff}^{\sigma}=\frac{\kappa_{ff}}{n_\sigma}F_\sigma
\end{equation}
where, $\sigma$ represents both magnetoionic modes and $\sigma=+1$ for o-mode and -1 for x-mode. In cold plasma approximation, $F_\sigma$ and $n_\sigma$ are defined as:
\begin{align}
    F_\sigma&=2\frac{\sigma\sqrt{P}[u\sin^2\theta+2(1-v)^2]-u^2\sin^2\theta}{\sigma\sqrt{P}[2(1-v)-u\sin^2\theta+\sigma\sqrt{P}]^2}\\
    n_\sigma^2&=1-\frac{2v(1-v)}{2(1-v)-u\sin^2\theta+\sigma\sqrt{P}}
\end{align}
where, $u=\nu_B^2/\nu^2$, $v=\nu_p^2/\nu^2$, $P=u^2\sin^4\theta+4u(1-v)^2\cos^2\theta$, $\nu_B$ is the gyro frequency, $\nu_p$ is the electron plasma frequency, $\theta$ is the angle between the magnetic field ($\mathbf{B}$) and the LoS. However, under quasi-longitudinal (QL) approximation, one can write:
\begin{equation}
    \kappa_{ff}^\sigma=\frac{\kappa_{ff}}{(1+\sigma\sqrt{u}\cos\theta)^2}
\end{equation}
Hence, the radiative transfer equation can be written as:
\begin{align}
    T_B^\sigma&=\int T(\tau^\sigma)e^{-\tau^\sigma}d\tau^\sigma\\
    \tau^\sigma&=\int\kappa_{ff}^\sigma(l)dl
\end{align}
where $T_B^\sigma$ and $\tau^\sigma$ represent the brightness temperature and the optical depth of the two modes, respectively. Due to the difference in their absorption coefficients and their sense of polarization being opposite, circular polarization is generated. Let us denote the degree of polarization as $V_{frac}$. Then, for optically thin regions, we can write:
\begin{equation}
    V_{frac}=\frac{V}{I}=\frac{T_B^R-T_B^L}{T_B^R+T_B^L}=\frac{T_B^x-T_B^o}{T_B^x+T_B^o}=\frac{\kappa_{ff}^x-\kappa_{ff}^o}{\kappa_{ff}^x+\kappa_{ff}^o}\approx2\sqrt{u}\cos\theta
\end{equation}
Substituting u and calculating the numerical values gives:
\begin{equation}
    B\cos\theta\text{ [G]}\approx5400\frac{V_{frac}}{\lambda\text{ [cm]}}
\end{equation}
This expression gives a very convenient way to relate the observed $V_{frac}$ to the LoS component of $\mathbf{B}$. The ability to measure very low $V_{frac}$ over a large range of frequencies will allow SKA telescopes to measure $B_{LoS}$ in the corona through the continuum thermal free-free emission. In the following subsection, we discuss a few studies that have used such methods to estimate the coronal magnetic field.
\subsubsection{Magnetic field measurements}
Several studies have shown that thermal free–free emission—particularly its weak circular polarization—can provide quantitative constraints on LoS coronal magnetic fields. The classical work of \citet{bogod1980measurements} demonstrated that the opacity difference in the o-x mode can produce reliable estimates of ($B_\parallel$), recovering a value of 40 G above a plage. Building on this, later articles formalized the inversion and clarified its limits, showing that the weak-field linear approximation is accurate when ($\nu_{B}/\nu\ll1$) and when independent temperature information is available. Modern analyses, such as \citet{iwai2014coronal}, combined the  Nobeyama Radioheliograph \citep[NoRH;][]{NoRH_Nakajima_1994} Stokes (V) maps with EUV differential emission measure (DEM) constraints using the observations of the Atmospheric Imaging Assembly \citep[AIA;][]{AIA_Lemen2012} onboard Solar Dynamics Observatory (SDO) and retrieved coronal fields of $\sim84$ G in post-flare loops.
\citet{sastry_circular} showed that even very weak coronal magnetic fields ($\sim$0.2-1 G) at heliocentric distances of 1.5–3 $R_\odot$ produce measurable circular polarization in low-frequency (10–80 MHz) thermal free–free emission. Recent studies by \citet{ramesh2010estimation} and \citet{mccauley2019low} have estimated circular polarization of 10-15\% and 5-8\% for coronal streamers and coronal holes, respectively. In the millimeter/submillimeter regime, circular polarization measurements of the Sun can enable diagnostics of the chromospheric LoS magnetic field via weak Stokes V signals from thermal free–free emission. Initial ALMA (Atacama Large Millimeter/submillimeter Array) solar polarimetry efforts have reported detectable circular polarization in active regions \citep{shimojo2024observing}, while simulations and observational analyses show that ALMA-band measurements can reliably recover chromospheric magnetic fields from such signals\citep[e.g.,][]{loukitcheva2017millimeter,loukitcheva2020measuring}.
However, estimating the magnetic field strength across a range of coronal heights through faint free-free circular emission requires wide-bandwidth sensitive radio observations. We discuss this in later sections how such needs will be fulfilled by SKA telescopes in section \ref{sec:role_of_ska}.

\subsection{Gyro-resonance}
\subsubsection{Emission mechanism}
In the early 1960s, it was independently demonstrated theoretically by \citet{zheleznyakov1962origin} and \citet{kakinuma1962model} that gyroresonance emission dominates centimetric emission from active regions. The million kelvin solar corona can be characterized by two crucial frequencies -- the plasma frequency ($\nu_p=8980\sqrt{n_e}$ Hz where $n_e$ is electron density in $cm^{-3}$) and the gyro frequency ($\nu_B=2.80\times10^6B$ Hz where B is measured in Gauss) due to the magnetic Lorentz force. The gyrating electrons can emit and absorb radiation at the fundamental and low harmonics of the gyro frequency ($\nu=s\nu_B$ where s=2,3,4...). Hence, the gyroresonance emission comes from a very narrow iso-Gauss layer in the corona. 
Electromagnetic waves propagating through coronal plasma have two circularly polarized modes \citep{ginzburg1962propagation} under most conditions. Of them, the x-mode gyrates in the same sense as the electron in the $\mathbf{B}$ field, and the o-mode gyrates in the opposite sense. Hence, the two modes interact with the thermal electrons differently, where the x-mode interacts strongly, compared to the o-mode. The equation for optical depth of gyroresonance for a thermal (Maxwellian) velocity distribution of the electrons has been derived and discussed in many articles \citep[e.g.][]{zheleznyakov1970radio,dulk_ff}. A simplified expression was provided by \citet{white1997radio} :
\begin{equation}
    \tau_{x,o}(s,\nu,\theta)\approx0.0133\frac{n_es^2}{\nu s!}\left(\frac{s^2\sin^2\theta}{2\mu}\right)^{s-1}L_BF_{x,o}(\theta)
\end{equation}
where, $\mu=m_ec^2/k_BT$, the scale length $L_B=B/\nabla B$ is calculated along LoS and the function $F_{x,o}(\theta)$ is defined such that it is unity at $\theta=90^o$ for the x-mode. For angles far away from $90^o$, one can approximate, $F_{x,o}(\theta)\approx(1\pm\cos\theta)^2$ while the `+' sign corresponds to x-mode and the `-' sign corresponds to the o-mode.

It can be understood that if the photospheric magnetic field is sufficiently low or the frequency is sufficiently high, no gyro-resonance emission will be seen. This is expected because the third and the second harmonic layers will be below the transition region (TR), making it impossible for these frequencies to escape the TR. When the frequency is appropriately low and/or the magnetic field is sufficiently high, the third harmonic enters the transition region, and a strong (almost 100\%) circular polarization in the sense of x-mode is observed \citep{shibasaki1994purely,nindos2000observations}. Gyro-resonance emission for fourth harmonics and higher do not have significant optical depth in the corona and hence are not usually important.
Although by definition gyro-resonance radiation is emitted in discrete frequencies, due to the variation of the magnetic field with height, the resultant emission appears continuous in nature. But this is only true if the magnetic field decreases monotonically with height. However, there are exceptions to these. For example, when there's a hot structure in corona, the frequency of the third harmonic of the emission will be higher than in nearby regions. This gives rise to cyclotron lines, the width of which depends on the magnetic field gradient. \citet{zhelezniakov_and_zlotnik} showed that the bandwidth of these lines can be written as:
\begin{equation}
    \frac{\delta\nu}{\nu}=\sqrt{2}\beta_T\cos{\alpha}
\end{equation}
where $\delta \nu/\nu$ is the fractional bandwidth, $\beta_T$ is the ratio of thermal velocity of electrons and the speed of light, and $\alpha$ is the angle between the magnetic field and the LoS.
The following subsection summarizes selected studies that have utilized these methods to infer the coronal magnetic field.
\subsubsection{Magnetic field measurements}
Since the gyro-resonance (g-r) emission and the magnetic field are closely related, it is a valuable tool in studying the atmospheric layers above active regions. 
With early modelling and imaging \citet{alissandrakis1980model} established it as a direct probe of strong ($\approx 10^2-10^3$ G) magnetic layers above sunspots. The g-r opacity depends strongly on electron temperature and density, harmonic number, wave mode, and the angle between $\mathbf{B}$ field and LoS (x-mode ≫ o-mode; 2nd harmonic ≫ 3rd in opacity) \citep{akhmedov_1982,shibasaki1994purely}. High-resolution, multi-frequency imaging and spectral modeling have since refined this technique and been reviewed in comprehensive syntheses \citep{white2004coronal,lee2007radio}, while more recent observational and data-driven modeling studies \citep[RATAN-600/SSRT/NoRH;][]{stupishin2018modeling,nita2018dressing,alissandrakis2019modeling}and updated theoretical treatments of thermal gyro-resonance/free-free emission \citep[e.g.][]{fleishman_ff} have improved the mapping of harmonic layer heights and enabled routine magnetic-field estimates from the spectral onset of polarized emission (routinely applied to RATAN-600 and other data). Contemporary efforts focus on combining NLFFF / DEM-driven 3D models with multi-frequency imaging to produce full-Stokes synthetic maps and quantitative magnetic diagnostics.

\begin{figure}[!h]
    \centering
    \includegraphics[width=0.8\linewidth]{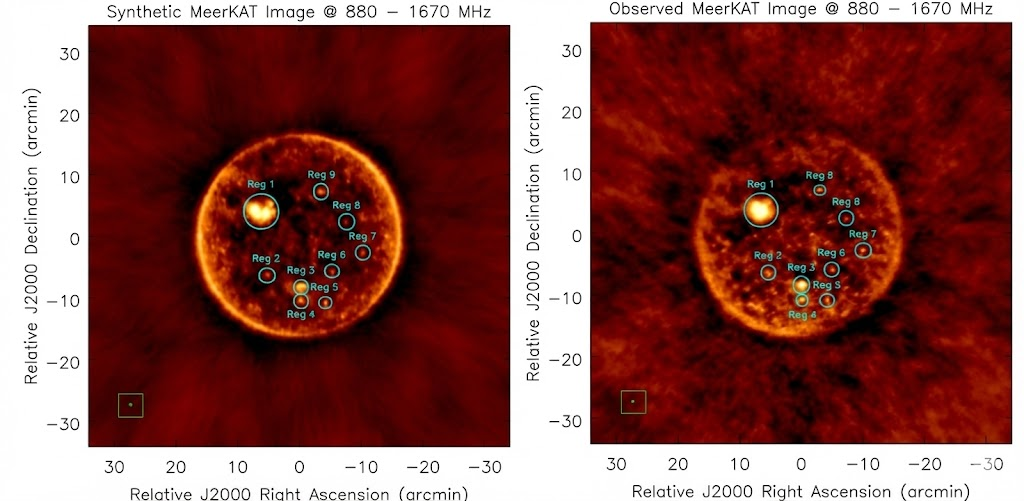}
    \caption{Comparison between synthetic and observed MeerKAT radio images on 2020 September 27, 10:45 UTC. Left panel: synthetic MeerKAT solar radio image.
Right panel: observed MeerKAT solar radio image. In both images, multiple bright
regions have been detected. Some of them are marked by cyan circles. Small green-filled circles at the bottom left corner marked by a green box represent the PSF of
the images. Figure adapted from \citet{Kansabanik_2024_meerKAT}}
    \label{fig:bright_points}
\end{figure}

Another interesting application can be studying coronal bright points which are a class of coronal structures which practically make them behave like mini-active regions. Like active regions, they are also quite stable structures and can last for several days. Their magnetic fields strengths lie in the range 10--200 $G$, higher than the diffuse quiet solar corona, but lower than that generally seen in active regions. This suggests that it might be possible to detect the gyro-resonance emission from coronal bright points, albeit weaker and at frequencies lower than that generally observed in active regions. \citet{Kansabanik_2024_meerKAT,kansabanik_2025_meerKAT} have already demonstrated that with the availability of high fidelity solar radio imaging using MeerKAT, coronal bright points can be detected quite easily all over the solar disc (see figure \ref{fig:bright_points}). With its broad frequency coverage and enhanced sensitivity, the SKA telescopes are expected to not only detect coronal bright points, but also enable modeling their spectrum and determine their magnetic field. Additionally, since coronal bright points are present in much larger numbers compared to active regions, coronal magnetic field measurements might become available over much larger areas once this method becomes available.

\subsection{Gyrosynchrotron Emission}
\subsubsection{Emission Mechanism}
Mildly relativistic electrons trapped in flaring coronal loops produce gyrosynchrotron (GS) radiation. The emission depends sensitively on the magnetic field magnitude (B), pitch-angle distribution, electron energy spectrum, and the viewing geometry. 
The classical expressions for emission and absorption of gyro-synchrotron emission in a magneto-active plasma were given by \citet{ramaty_gs_1969}. \citet{dulk_ff} gave the expected spectral shape for both thermal and power law non-thermal electron distributions. The emission is quasi-continuous in frequency and has a peak at the low harmonics of the gyrofrequency. However, magnetic fields in flaring sources are highly inhomogeneous, which makes direct magnetic field measurement less straightforward than other methods. As discussed in the following text, most studies assume a homogeneous distribution of energetic electrons. However, simplistic models do not agree well with observational data \citep{Kansabanik2024_gs_StokesV}. Several other studies have considered inhomogeneous and anisotropic distributions of non-thermal electrons and temporal variability \citep[e.g][]{fleishmann_2003,tzatzakis_2008,simoes_2020,nita_2015}.
\subsubsection{Magnetic Field Measurements}
\label{gyrosynchrotron_mag_field_measurements}
Despite difficulties, earlier studies included detailed modelling of flaring loop observations at high radio frequencies. \citet{nindos2000observations} derived magnetic-field strengths of 870 G at the footpoints and 270 G at the loop top of a flaring structure by combining data from Karl G. Jansky Very Large Array \citep[VLA;][]{Perley2011_vla} and spectral data from Owens Valley Radio Observatory \citep[OVRO;][]{ovro_lwa_Hallinan_2023}. Comparable values, spanning 1,700–200 G, were later obtained from Nobeyama \citep{Nakajima1985Nobeyama} 17 and 34 GHz imaging by Kundu et al. (2001, 2004), \citet{tzatzakis_2008}, and \citet{kuznetsov2015spatially}. In recent years, with the Expanded Owens Valley Solar Array \citep[EOVSA;][]{Gary2018_EOVSA} providing spatially resolved, high-cadence broadband spectra, direct spectral fitting of GS emission has been done. One of EOVSA’s earliest limb-flare studies, involving imaging at 30 frequencies between 3.4–18 GHz, yielded preliminary magnetic-field estimates of 150–520 G \citep{gary2018microwave}. \citet{fleishman2020decay} showed that the decrease in stored magnetic energy is sufficient to power solar flares. \citet{chen2020measurement} measured the magnetic field profile along the reconnecting current sheet and compared it with MHD simulations.
\begin{figure}[!h]
    \centering
    \includegraphics[width=0.7\linewidth]{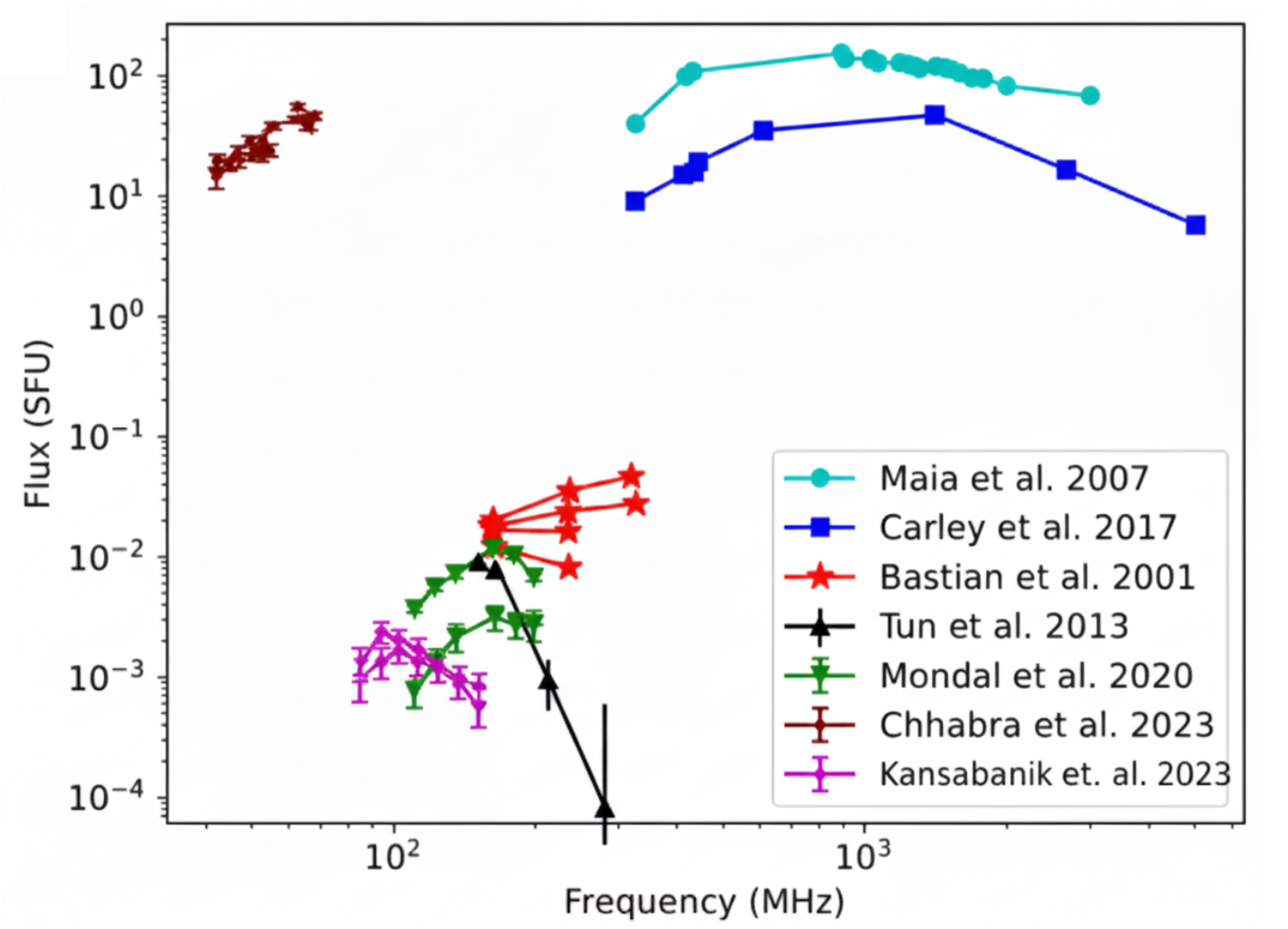}
    \caption{Comparison of gyrosynchrotron emission spectra from CME plasma as reported in various studies. The figure is adapted from \cite{Kansabanik2023_CME1}}
    \label{fig:gs_spectra}
\end{figure}
Gyrosynchrotron diagnostics have also been applied to Coronal Mass Ejections (CMEs), provided that the associated type IV emission arises from GS rather than plasma emission processes—a distinction identifiable from low brightness temperatures and peaked spectra \citep{Klein_Trottet1984}. Although radio-bright CMEs are rare, early work by \citet{Gopalswamy_kundu1987} estimated about 2.5 G at $\sim2.3 R_\odot$, and subsequent studies \citep[e.g.,][]{bastian2001, Maia2007,TunVourlidas2013,Bain2014,Carley2017,Mondal2020a, Kansabanik2023_CME1} reported values ranging from 0.3 to 23 G at 1.3–2.7 $R_\odot$, with the spread reflecting intrinsic differences among CMEs rather than ambient coronal conditions. Notably, all of these studies employed homogeneous source models and simplified emissivity expressions, which could not provide enough constraints. These underscore the need for more sophisticated modelling in future CME magnetography.
In order to achieve reliable diagnostics of GS emission, one needs highly sensitive data with high spectral, spatial and temporal resolution. With SKA telescopes, such needs will be fulfilled. We will return to this discussion in Section \ref{sec:role_of_ska}.

\subsection{Plasma emission}
\label{subsec:plasma-emission}
\subsubsection{Emission Mechanism}\label{plamsa_emission_mechanism}
Plasma emission is a coherent emission mechanism produced by non-thermal electrons -- usually accelerated by magnetic reconnection or MHD shocks -- that excite Langmuir waves in the background coronal plasma as they propagate \citep{Melrose1977}. These electrostatic plasma waves convert into the escaping electromagnetic radiation through non-linear processes near the local plasma frequency and its harmonics. The fundamental emission (F) occurs when Langmuir waves are scattered by plasma ions or otherwise decay, whereas the harmonic emission (H) arises from the coalescence of two Langmuir waves. The background magnetic field influences both the nonlinear conversion processes and the properties of the escaping magnetoionic modes (x and o modes), thereby determining the total degree of polarization \citep{MelroseSy1972, MelroseDulkSmerd1978}. Most solar radio bursts (such as type I, II, III, V, and some IV bursts) at decimetric and higher wavelengths are produced by plasma emission. For a detailed discussion on the spectroscopic properties of solar radio bursts, refer to the chapter on "Solar Radio Bursts" by Kumari et al.

Many theoretical studies have examined the expected polarization properties of plasma emission. \citet{MelroseSy1972} examined this regime assuming weak magnetic fields and Langmuir-wave phase speeds well below the speed of light. 
They found that a one-dimensional Langmuir-wave distribution aligned with the magnetic field favors o-mode in the F component. Hence, when the emission frequency lies below x-mode cutoff ($\omega<\omega_x$), the F component shows circular polarization arising from the o-mode.
Later, with extended treatment using isotropic Langmuir-wave distributions and narrow angular ranges, \citet{Dulk1976} and \citet{Melrose1977} predicted that 
the H component becomes partially polarized in the x- mode. 

Several studies further demonstrated that H emission can appear in the o-mode only if the Langmuir-wave propagation is tightly confined within a small cone -- roughly $\theta \lesssim 20^\circ$ -- around the magnetic field direction, a condition requiring relatively strong magnetic fields \citep{MelroseSy1972, MelroseDulkSmerd1978,  Melrose1980,Zlotnik1981}. When this confinement is absent, partial x-mode polarization is expected. Because the degree of o-mode polarization (dcp) depends sensitively on the field strength, H polarization provides a potential diagnostic for the coronal magnetic field as follows:
\begin{equation}
    \mathrm{dcp} = \frac{11}{48}\left( \frac{\nu_{B}}{\nu_{p}} \right) \, \!\left| \cos \theta \right|
    \label{equ:dcp_to_B}
\end{equation}

where $\nu_B=2.80\times10^6B$ Hz (B is the background magnetic field strength measured in Gauss) is the gyrofrequency, $\nu_p$ is the local plasma frequency, and $\theta$ is the angle between the magnetic field ($\mathbf{B}$) and the direction of propagation of radiation ($\mathbf{k}$). 

Since the polarization properties of the radio bursts are directly related to the local magnetic field strength and orientation, they are one of the most direct remote-sensing probes for the magnetic field estimation from the lower corona upto the inner heliosphere.
However, the propagation effects such as scattering, mode-coupling, as will be discussed in section \ref{sec:propagation_effects}, change the polarization signatures, making it hard to interpret. Despite these challenges, there have been efforts in constraining the magnetic field from various radio bursts. In the following paragraphs, we discuss how polarimetric properties of radio bursts, particularly their Stokes-V signatures, have been used to infer coronal magnetic-field strength and topology.

\subsubsection{Magnetic field measurements}
\label{plasm_emission_mag_measurements}
\begin{enumerate}
    \item \textbf{Type II bursts:} Type II bursts are produced by shocks that are often driven by flares and CMEs. The favorable conditions for the occurrence of a type II emission are believed to be shock regions with a quasi-perpendicular geometry relative to the ambient coronal magnetic field \citep{Cho2011,Zimovets2015,Zucca2018,Kouloumvakos2021}.
    Type IIs occur over a wide frequency range, from 100s of MHz to a few kHz, allowing them to probe magnetic field conditions across a wide range of height in solar atmosphere.
    In dynamic spectrum, type II bursts appear as slowly drifting emissions at the fundamental and harmonic of the local plasma density. Each of these emission bands is frequently observed to exhibit band-splitting, which is commonly interpreted as radiation originating from the upstream and downstream regions of the shock front \citep{Smerd1974, Smerd1975, vrsnak2001,Zimovets2012, Zucca2018}. However, an alternative explanation for band-splitting exists, which interprets the split bands to originate from different parts of the shock front. \citep{Du2014, Zimovets2015, Bhunia2023, Zucca2025}. 
    
    Type II bursts, exhibiting split bands, are used to estimate the shock-entrained magnetic field strength under the assumption that the split bands originate from the upstream-downstream regions of the shock front \citep{Smerd1974, Smerd1975}. Following the Rankine-Hugoniot relation, the Alfv\'en mach number ($M_A$) is related to the shock compression ratio (X). For a quasi-perpendicular shock approximation and plasma $\beta \ll 1$, the relation can be written as \citep{Vrsnak2002},
    \begin{equation}
        M_A = \sqrt{\frac{X (X + 5)}{2 (4 - X)}}
    \end{equation}
    where the compression (X) is defined as, 
    \begin{equation}
        X = \frac{N_{e2}}{N_{e1}}
      = \left( \frac{f_2}{f_1} \right)^2
    \end{equation}
    here $N_{e2}$ and $N_{e1}$ are the downstream and upstream regions of the shock and $f_2$ and $f_1$ are the corresponding plasma frequencies, i.e., the frequencies of the split bands. 

    Finally, from $M_A$ ($= v/v_A$, where $v$ is the shock speed and $v_A$ is the Alfv\'en speed), one can estimate $v_A$ and thus the magnetic field strength from an emission with frequency of $f$ as,  
    \begin{equation}
        B\;[\mathrm{G}] = 5.1 \times 10^{-5} \; f\;[\mathrm{MHz}] \; \times \; v_A\;[\mathrm{km\,s^{-1}}]
    \end{equation}

    Many studies have employed this approach to estimate the magnetic field strength, combining the drift rate (shock speed), coronal density model (height of the emission), and the band-split ratio (shock compression) \citep{Vrsnak2002, Zimovets2012, Hariharan2015,Mahrous2018, Kumari2017a, Kumari2019}. Although this method has been widely used for metric and interplanetary type IIs with band-splitting, its approach relies heavily on assumed electron density models, shock geometry, and the interpretation of the split bands. 
    Several later studies have included multi-wavelength information (whitelight coronagraph, EUV images) in an attempt to reduce these model-dependent uncertainties \citep{Gopalswamy2012, Kumari2017typeII_band, Kumari2019}. 
    
    In the literature, type II bursts are reported to show a diverse range of circular polarization properties, from weak or unpolarized to strongly polarized \citep[e.g.,][]{Komesaroff1958, Hariharan2014, Ramesh_2022, Alissandrakis2021}. However, the fine structures, known as herringbones, exhibit a higher polarization fraction than the burst continuum. The degree of circular polarization, in the sense of o-mode, from harmonic emission has also been used to estimate the magnetic field strength through equation \ref{equ:dcp_to_B} \citep{Hariharan2014, Kumari2017typeII_band, Ramesh_and_Kathiravan2022, Ramesh_2023}.

    \item \textbf{Type III bursts:} 
    Type III radio bursts are produced by beams of energetic electrons -- typically accelerated during magnetic reconnection -- that propagate along open magnetic field lines from the low corona into interplanetary space. In dynamic spectra, they appear as intense, rapidly drifting features and are detected over an exceptionally broad frequency range, from GHz to kHz, making them excellent tracers of open magnetic structures throughout the corona and heliosphere.

    The polarization properties of Type III bursts offer important diagnostic information. As discussed in Section~\ref{plamsa_emission_mechanism}, the sense and degree of circular polarization of the F and H components can be used to infer both the magnetic-field orientation and, under appropriate conditions, its strength \citep[e.g.][]{MelroseSy1972, MelroseDulkSmerd1978, DulkSuzuki1980, Zlotnik1981}. Although theory predicts that the fundamental component can be nearly 100$\%$ circularly polarized, observations typically show much lower values due to the propagation effects from the intervening coronal medium. In their survey of 997 bursts, \citet{DulkSuzuki1980} reported average degrees of circular polarization of $\sim35\%$ for F, $\sim11\%$ for H, and only $\sim6\%$ for structureless events. At higher, near-microwave frequencies, events with polarization approaching 100$\%$ have been documented \citep{Wang2003}.

    Observations across multiple instruments and frequencies further support the diagnostic potential of polarization. Several studies have shown that the degree of circular polarization tends to increase with observing frequency  \citep{BenzZlobec1978, Mercier1990}. The temporal evolution of polarization is also revealing: in many events, the polarization peaks before the total intensity, suggesting rapid variations in either the emission conditions or propagation effects such as mode coupling and scattering \citep{Benz1982, Mercier1990}. These trends can help distinguish whether the radiation originates from a single magnetic-polarity region or involves a transition between fundamental and harmonic emission.

    Polarimetric imaging studies remain rare
    , largely due to calibration challenges at low radio frequencies. A significant advancement was made by \citet{Rahman2020}, who used the Murchison Widefield Array \citep[MWA;][]{Tingay2013_MWA} to obtain Stokes I and V images of Type III bursts between 80 and 240 MHz. They found that the circular polarization fraction increases with frequency and is strongest near the onset of each burst -- consistent with expectations that the fundamental component appears first. As the bursts evolved, the polarization systematically decreased, which they attributed to enhanced scattering and propagation-induced depolarization. Although plasma emission is intrinsically expected to be circularly polarized, in a recent polarimetric imaging study, \citet{Dey2025linear_pol} has demonstrated a robust detection of linear polarization associated with a Type III burst. They interpret it as a propagation-induced effect rather than an intrinsic property of the emission. These results highlight the capability of modern low-frequency interferometers to investigate the magnetic environment of type III-producing electron beams. 

    \item \textbf{Type IV bursts:}
    Type IV bursts are long-duration broadband continuum emission associated with flares and the lift-off and the wakes of CMEs. They span a broad frequency range—from decimetric to decametric wavelengths—and often persist for tens of minutes to hours \citep[e.g.,][]{wild1972radio,Salas_and_ludwig2020}. Based on their spectral morphology, type IV bursts are commonly divided into two subclasses \citep{Boischot1957}. Moving type IV (IVm) bursts exhibit an outward propagation of the source region, producing a characteristic frequency drift that typically tracks the CME core or flux rope. In contrast, stationary type IV (IVs) bursts appear at fixed coronal heights, often above post-flare loops, and show little to no systematic drift \citep{kundu1965}.

    The emission mechanisms responsible for type IV bursts remain a subject of active debate. Depending on the coronal environment and electron population, type IV emission has been interpreted as gyrosynchrotron radiation from mildly relativistic electrons \citep[e.g.][]{kai1969, Carley2017}, plasma emission driven by coherent processes \citep[e.g.][]{Duncan1981, hariharan2016}, or, in rare cases, electron–cyclotron maser (ECM)  \citep[e.g.][]{MelroseDulk1982,vasanth2016}. Observations further indicate that the emission mechanism can evolve during the course of an event, reflecting changes in magnetic topology, density, and electron acceleration conditions \citep{morosan2019}.

    While the sense and degree of circular polarization can, in principle, be used to estimate magnetic-field strength in type IV sources produced by plasma emission, only a limited number of studies have explored the polarimetric properties of such events in detail. By contrast, a substantial amount of work has used gyrosynchrotron type IV bursts. Particularly the IVm sources associated with CMEs are used to estimate the magnetic field entrained within CME flux ropes. These diagnostic applications have been discussed in Section \ref{gyrosynchrotron_mag_field_measurements}.
\subsection{Fibre and zebra bursts}
Fibre bursts and zebra pattern bursts are fine structures of broadband type IV solar radio bursts. They can serve as valuable diagnostics of the coronal magnetic field. Fibre bursts are narrowband features with intermediate frequency drift rates, and both whistler-wave and Alfvén-wave models have been proposed for their origin \citep{kuijpers1975generation,bernold1983fiber, benz1998intermediate}. The frequency drift rate of fibre bursts is directly related to magnetic field strength, allowing the field in the flaring source region to be derived. This approach has been validated observationally, with fibre burst field estimates agreeing with extrapolated potential-field models to within a factor of $\sim$0.6–1.4, supporting their use as a 3D coronal magnetic field probe in post-flare loops \citep{aurass2005fiber}.\\ 
Zebra patterns, most commonly explained via the double plasma resonance (DPR) mechanism have been analysed alongside type II band-splitting to constrain the coronal magnetic field at heights of $\sim$1.9–2 solar radii, yielding field strength estimates on the order of $\sim$0.43 G \citep{stanislavsky2015coronal}. At lower coronal heights, zebra patterns and co-temporal fibre bursts have been used together as complementary diagnostics providing collateral evidence for field strengths in zebra pattern source regions \citep{tan2012microwave}.

\end{enumerate}

\begin{figure}[!h]
    \centering
    \includegraphics[width=0.99\linewidth]{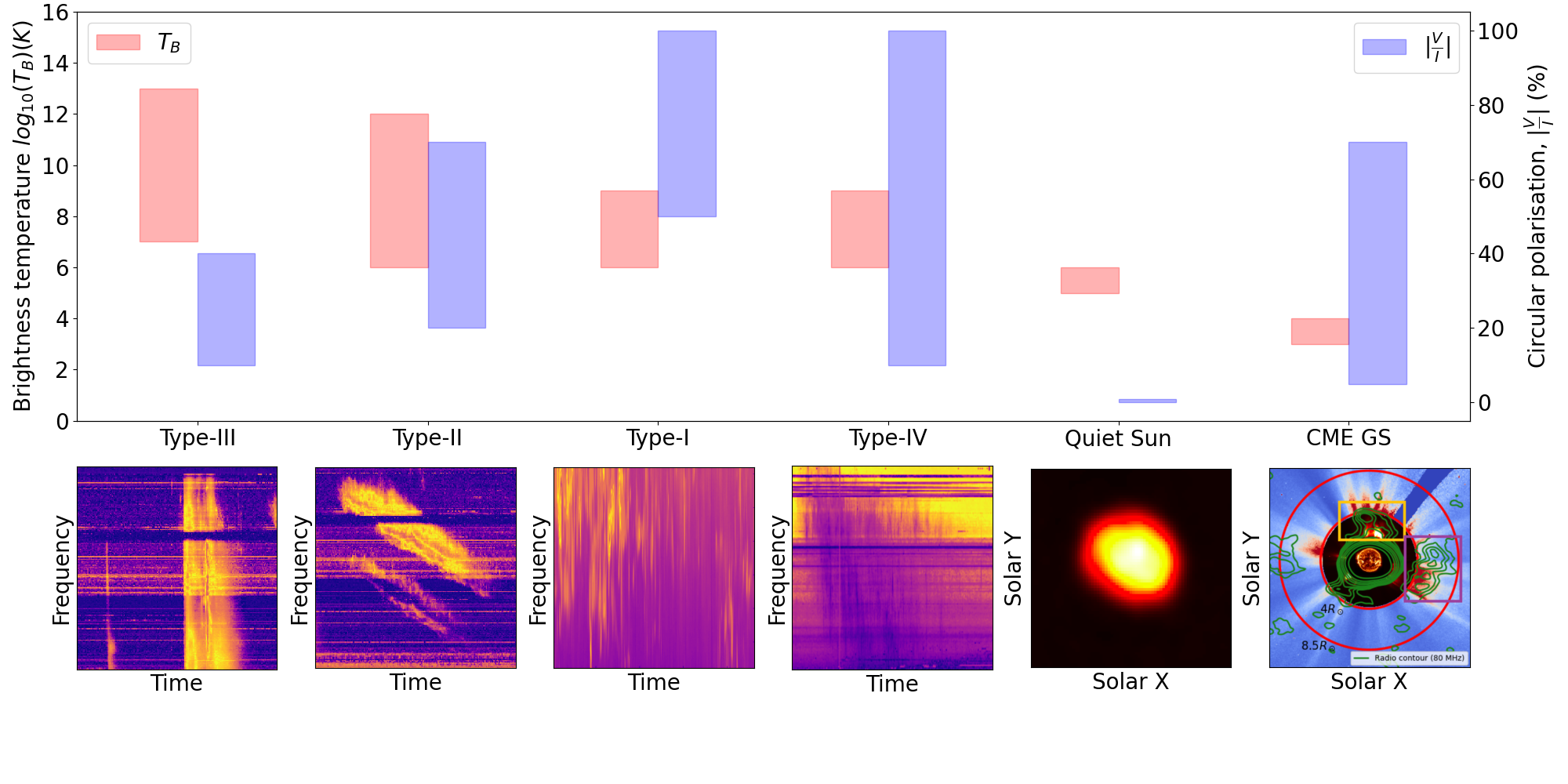}
    \caption{The upper panel illustrates the expected brightness temperatures (red bars) and circular polarization fractions (blue bars) of several low-frequency solar radio emissions, including type III, II, I, and IV bursts, quiet-Sun free–free emission, and CME-associated gyrosynchrotron radiation (from left to right). The lower panel presents corresponding dynamic spectra and radio images. The type II–IV spectra are derived from the Learmonth spectrograph data, while the type I dynamic spectrum and the quiet-Sun and CME images are obtained from MWA observations. The dynamic spectra highlight the wide variety of spectral and temporal behaviors across different burst types, whereas the radio images emphasize the broad range of spatial structures that give rise to these emissions. This figure is adapted from \cite{kansabanik2022working}.  }
    \label{fig:emission_tb_pol}
\end{figure}

\subsection{Propagation effects}
\label{sec:propagation_effects}
During propagation through the coronal plasma, the polarization of the electromagnetic field changes along the path. As a result the observed polarization is not the same as that at the origin of the emission. In a few special cases, the circular polarization of the wave changes sign due to the orientation of the local magnetic field and the direction of propagation. The effect of polarization inversion in microwave sources was discovered by \citet{piddington1951solar}. It was later explained by \citet{cohen1960magnetoionic} as being the result of quasi-transverse (QT) propagation of microwaves. In this section, we describe how information about the magnetic field above the active regions can be extracted by using this effect.
\subsubsection{Mode coupling}
\begin{figure}[!h]
    \centering
    \includegraphics[width=\linewidth]{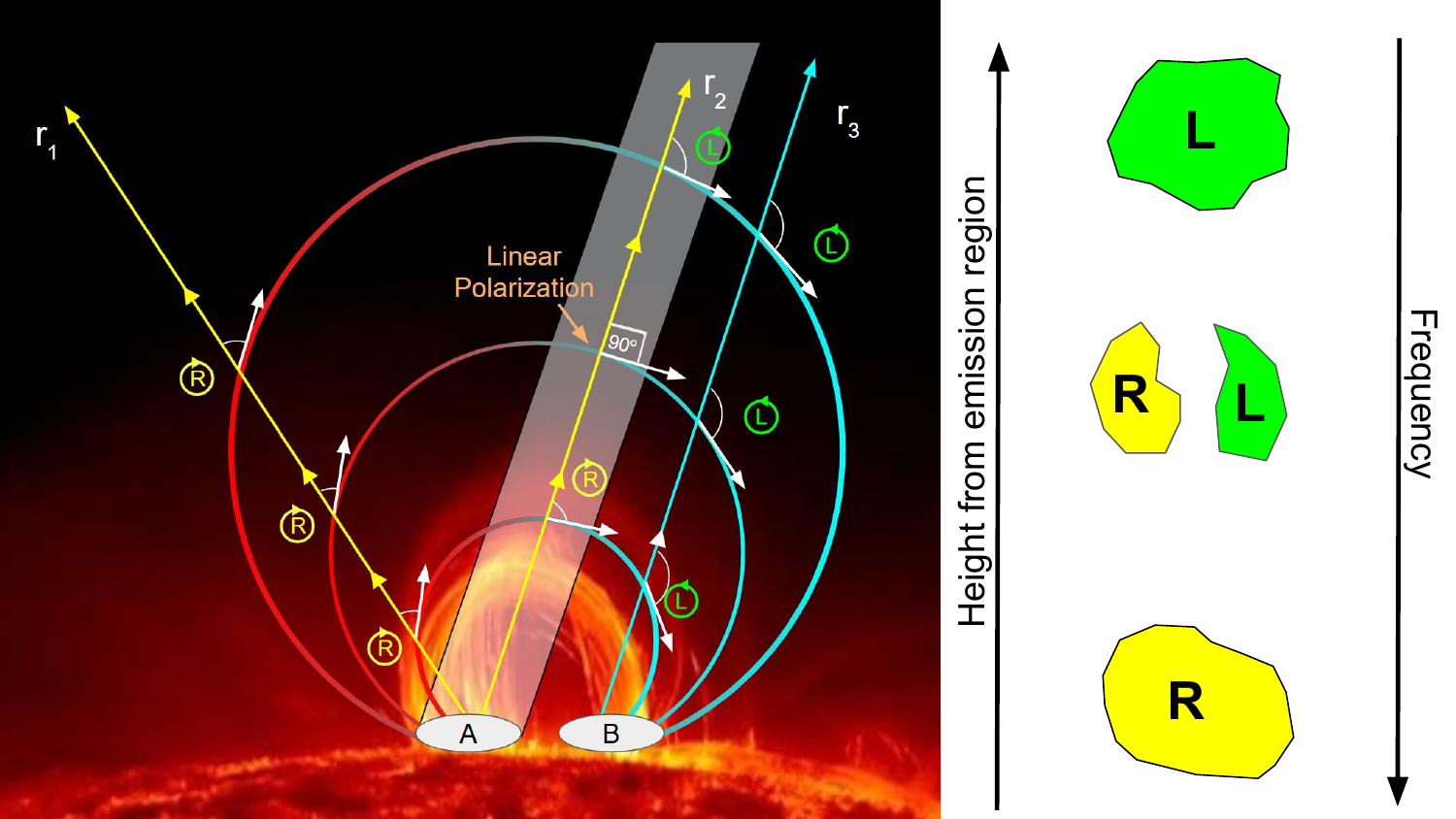}
    \caption{(Right) This schematic diagram shows propagation of x-mode through a bipolar magnetic field and it effects on the observed polarization depending on the viewing angle (R and L denotes RCP and LCP respectively). When it crosses a transverse magnetic field ($Ar_2$) the sense of circular polarization changes. Also in case of acute angles the polarization of x-mode is RCP and if the angle becomes obtuse it changes to LCP. The opposite happens for o-mode. (Left) This shows how the emitting region A appears in Stokes V radio images with frequency if the direction of propagation follows the grey region (figure inspired from \citet{bandiera1982diagnostic})}
    \label{fig:inversion}
\end{figure}

The EM wave passing through the cool and collisionless plasma (e.g. the solar corona) can be expressed as a linear combination of two opposite elliptically polarized modes (x-mode and o-mode). The ellipticity of the polarization of these two modes depends on the angle $\theta$ between the magnetic field vector ($\mathbf{B}$) and the direction of propagation ($\mathbf{k}$). For most values of $\theta$, these two modes are circularly polarized (quasi-longitudinal propagation or QL). At $\theta\approx90^{\circ}$, both of these modes become linearly polarized, with the o-mode aligning with the $\mathbf{B}$ and the x-mode being perpendicular to it (Quasi-transverse propagation or QT, \citet{zheleznyakov1977propagation}).\\
When an EM wave passes through a region of transverse magnetic field, depending on the strength of the magnetic field and the frequency of the wave, an inversion of circular polarization can occur (see figure \ref{fig:inversion}).
The wave-mode coupling theory discusses this interaction in detail. The parameter dictating the interaction known as the coupling parameter Q was derived by \citet{cohen1960magnetoionic} :
\begin{equation}
    Q=\frac{16\ \pi^2\ m_e^4\ c^4}{q_e^5}\ \frac{\nu^4\ |d\theta/ds|}{N_e\ B^3}
    \label{coupling}
\end{equation}
here, the constants have their usual meaning and $\nu$, $|d\theta/ds|$, $N_e$ and B are frequency, viewing angle variation scale length, electron density and magnetic field respectively. If $V_{in}$ is the initial degree of circular polarization before entering the quasi-transverse region (QTR) then the degree of circular polarization after crossing the region can be written as \citep{zheleznyakov1964polarization}:
\begin{equation}
    V_{obs}=V_{in}(2\exp(-\pi/2Q)-1)
\label{circular_eq}
\end{equation}
There can be 3 scenarios when the wave crosses the QTR:
\begin{itemize}
    \item $Q<<\pi/2$: Weak mode coupling; polarization changes sense ($V_{obs}\approx -V_{in}$)
    \item $Q=\pi/2\ln2$: Critical coupling; polarization becomes linear ($V_{obs}=0$)
    \item $Q>>\pi/2$: Strong coupling; polarization does not change sense ($V_{obs}=V_{in}$)
\end{itemize}
From \ref{circular_eq} we can write: 
\begin{equation}
Q=\frac{\pi}{2}\left[\ln\left(\frac{2}{V_{obs}/V_{in}+1}\right)\right]^{-1}
\label{coupling_2}
\end{equation}
 Using typical values of $N_e$ and $|d\theta/ds|$ in equation \ref{coupling}, one can simply write $Q=k\ \nu^4/B^3$ where k is constant\footnote{$k=\frac{16\ \pi^2\ m_e^4\ c^4}{q_e^5}\ \frac{\ |d\theta/ds|}{N_e}$}. Equating equation \ref{coupling} and equation \ref{coupling_2} one can derive expressions for B and $\nu$. The critical frequency at which the depolarization occurs can be written as:
\begin{equation}
    \nu_{crit}=\left(\frac{\pi\ \ln2}{2k}\right)^{1/4}B^{3/4}
\end{equation}
In a similar approach, one can also calculate the magnetic field from equation \ref{coupling_2}:
\begin{equation}
    B=\left(\frac{2k}{\pi}\right)^{1/3}\left[\ln\left(\frac{2}{V_{obs}/V_{in}+1}\right)\right]^{1/3}\nu^{4/3}
\end{equation}
\subsubsection{Limiting polarization}
The polarization of the two wave modes depend solely on the medium of propagation and not on the emission mechanism. In weak coupling the two wave modes are almost independent of each other. Hence, their polarization properties are affected by the local plasma conditions. However, when the waves encounter such a region where the coupling switches from weak to strong the polarization properties after that change no longer. This leads to the fact that the observed polarization properties can be significantly different from the source such as the case of magnetic field reversal or QT propagation.
For QL propagation, it can be shown from the coupling coefficient that strong coupling occurs at much lower values of electron density. Hence, in QL propagation the polarization properties are not affected much by the limiting polarization.
\subsubsection{Observations and Results}
The effect of circular polarization inversion has been observed for a long time \citep{kundu1965solar,zheleznyakov1970radio}. Due to the viewing angle geometry, the limb-ward part of an active region typically encounters the transverse field region (TFR), rather than the disk-ward part. In addition, the depolarization strip (the line that separates two opposite polarizations) deviates from the photospheric neutral line. The deviation in lower frequency (higher TFR), the displacement is larger than in higher frequency (lower TFR). Initial studies with low-resolution observation \citep{peterova1974influence} and later with high-resolution \citep{chiuderi1987microwave} have confirmed this effect. Under reasonable assumption of $N_e$, the magnetic field in the depolarization strip can be estimated \citep{kundu1984structure}. Table \ref{mag_measurements} summarises studies in constraining magnetic field above an active region at different heights.

\begin{table}[h]
    \centering
    \caption{Summary of magnetic field measurements from circular polarisation inversion. Adapted from \citet{alissandrakis2021radio}
    }
    \begin{tabular}{cccccc}
        \textbf{References} & \textbf{Frequency (GHz)} & \textbf{Height (Mm)} & \textbf{B (G)} \\
        \hline
        \citet{kundu1984structure} & 4.8 & 110 & 20 \\
        \hline
        \citet{alissandrakis1996coronal} & 4.8 & 130 & 10 \\
         & & 100 & 16 \\
         \hline
        \citet{segre2001evolution} & 4.8 & & 11.2–12.8 \\
        \hline
        \citet{gelfreikh1987measurements} & 7.5-15 & 120  & 16 \\
        \hline
        \citet{nagelis1992energetics} & 7.5-15 & 38 & 26 \\
        \hline
        \citet{lang1993magnetospheres} & 7.5-15 & 50–200 & 50–15 \\
         & & 200–300 & 10–5 \\
         \hline
        \citet{ryabov1999coronal} & 8.7-17 & 57–87 & 65–20 \\
         & 8.7-17 & 37–64 & 125–30 \\
         \hline
        \citet{ryabov2005coronal} & 5.7 & 50–90 & 30–10 \\
         & 17 & 15–38 & 110–50 \\
         \hline
    \end{tabular}
    \label{mag_measurements}
\end{table}
However, these studies did not have wide frequency coverage and were performed at a few discrete frequencies. With SKAO's enhanced sensitivity, the circular inversion can be observed with even finer spectral resolution, and the wide bandwidth will allow the study of multiple polarization inversions.
\section{The role of SKAO in coronal magnetography}
\label{sec:role_of_ska}

\subsection{Advancement in coronal polarimetric imaging with new generation instruments}

Legacy instruments like RATAN-600 have been used to study the coronal magnetic field for decades \citep[e.g.][]{bogod1980measurements}. While it lacks full two-dimensional imaging capability, it routinely measures Stokes V at microwave frequencies to estimate coronal magnetic field strengths above active regions through the gyroresonance emission mechanism. Along with RATAN-600, VLA \citep[e.g.][]{schmahl1982active,brosius2002measurements}, WSRT \citep{kundu1984structure} also helped in probing the solar corona and magnetic fields using microwave polarimetric imaging.
With the new generation of instruments — MWA, The LOw-Frequency ARray \citep[LOFAR;][]{lofar2013} and EOVSA — coronal magnetography through imaging studies of the quiet Sun and a wide range of active solar emissions has been demonstrated
across frequencies from a few megahertz to several gigahertz, as discussed in Section \ref{sec:mag_field_measurement}. More recently, the Mingantu Spectral Radioheliograph (MUSER; \citet{yan2021mingantu}) has begun spectropolarimetric imaging of the solar corona over a broad frequency range (0.4–15 GHz), providing a new observational capability for future coronal magnetography studies.
These efforts have shown the significant potential of low- and mid-frequency solar radio observations but have also highlighted key challenges that must be addressed for the SKAO. A detailed discussion on the challenges and requirements for facilitating solar science with SKAO can be found in the chapter by \citet{Oberoi01.2026.SKA}.

Reliable polarization calibration at low radio frequencies remains difficult due to strong direction-dependent instrumental effects, ionospheric phase distortions, and complex primary beam responses. For wide-field instruments such as MWA, LOFAR, The Owens Valley Radio Observatory Long Wavelength Array \citep[OVRO-LWA;][]{ovro_lwa_Hallinan_2023}, New Extension in Nançay
Upgrading LOFAR \citep[NenuFAR;][]{nenufar_Hicks2012}, and the upcoming SKA-Low, accurate modeling of the frequency and direction-dependent beam is particularly critical, as deviations between the modeled and true beam responses leave behind uncorrected
instrumental leakage. Such leakage results in polconversion, where Stokes I contaminates Stokes Q, U, and V. Errors in cross-hand phase calibration introduce polrotation, mixing the polarized Stokes parameters \citep{Hamakar1996, Sault1996}. 
These issues are further compounded by the scarcity of well-characterized polarized calibrators at low frequencies, making external polarization calibration difficult. In such cases, self-calibration approaches are often used. At higher frequencies, although the number of polarized calibrators increases, the intense solar flux often contaminates the calibrator scans, reducing their effectiveness.

Many of these calibration challenges have been systematically addressed by using the opportunity presented by the SKAO precursors and pathfinders. 
A dedicated automated polarimetric calibration and imaging pipeline that explicitly incorporates instrumental leakage has been developed for MWA \citep{paircars22a, paircars22b, paircars23}. A cross-hand phase estimation technique using the unpolarized sky has also been created and validated with MWA data, enabling more reliable polarization-angle measurements  \citep{cross_phase_Kansabanik2025}. LOFAR employs automated solar imaging pipelines capable of high-fidelity spectroscopic snapshot imaging \citep{Dey2025SIMPL}. Ongoing efforts aim to extend these pipelines to full-Stokes polarimetric calibration.

For the SKA-Mid precursor MeerKAT \citep{jonas2016meerkat}, the primary beam response has been measured holographically \citep{MeerKAT_primary_beam_Villier_2023}. Recent work has demonstrated broadband spectroscopic solar imaging with MeerKAT \citep{Kansabanik_2024_meerKAT, kansabanik_2025_meerKAT}, marking a major step toward full-Stokes solar polarimetry with the instrument, which remains an ongoing area of development.

Together, these efforts constitute crucial technical and algorithmic advances. The lessons learned from MWA, LOFAR, and MeerKAT can directly help with the calibration strategies, beam modeling requirements, and pipeline requirements for reliable solar observations with the SKA telescopes.

\subsection{Capabilities of SKA telescopes}
\begin{figure}[!h]
    \centering
    \includegraphics[width=\linewidth]{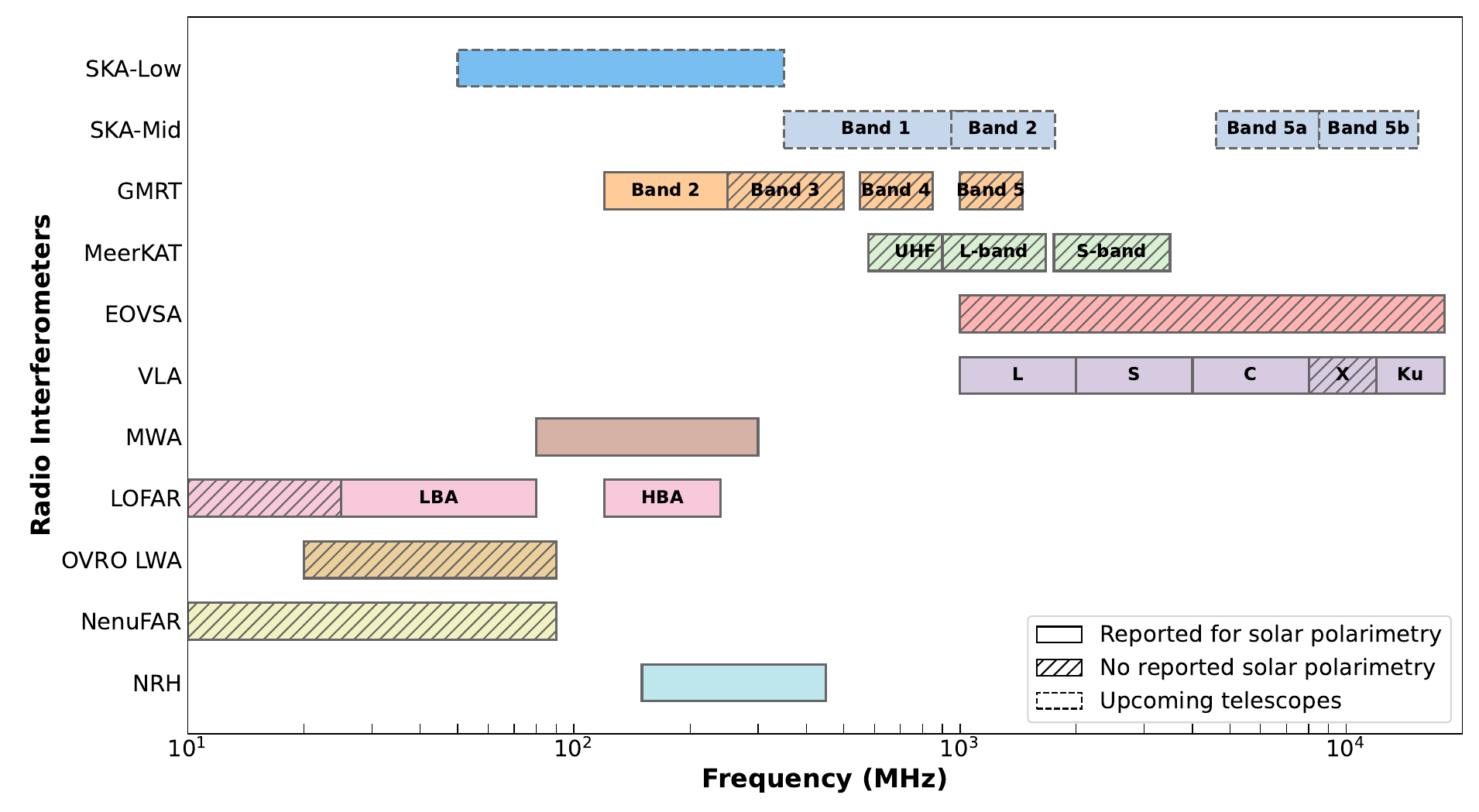}
    \caption{Bandwidth coverage of radio interferometers currently in operation for solar observations. Each interferometer is represented by its operational frequency bands, with solid colors indicating frequencies currently utilized for solar polarimetry studies and hatched patterns indicating frequency ranges where solar polarimetry observations have not been reported. The upcoming telescopes (SKA-mid and SKA-low) are indicated with dashed lines.}
    \label{fig:band_coverage}
\end{figure}

The upcoming SKA telescopes (SKA-Low and SKA-Mid) will provide unprecedented sensitivity and wide frequency coverage (Figure \ref{fig:band_coverage}), significantly advancing 
coronal magnetography and polarimetry. While the SKAO remains committed to achieving its full design baseline of 512 SKA-Low stations and 197 SKA-Mid dishes, initial operations will begin with an interim configuration known as AA* \citep{seethapuram_sridhar_2025}. This subsection refers to the AA* design of the SKAO.

SKA-Low consists of 307 stations, each containing 256 log-periodic dipole antennas arranged in a dense core with three spiral arms. This configuration provides extremely dense instantaneous monochromatic uv coverage and a maximum baseline of about 74 km. Similarly, SKA-Mid will comprise 144 dishes (80 new 15-m SKA dishes and 64 MeerKAT 13.5-m dishes), delivering dense uv coverage and maximum baselines up to $\sim$108 km (exclusion of SKA008 makes the maximum baseline 36km). Both arrays will produce high-fidelity spectro-polarimetric images with a much higher dynamic range
than the current instruments, along with excellent angular resolution.

As shown in Fig. \ref{fig:emission_tb_pol}, solar emissions span a wide range of brightness temperatures and polarization states. SKA’s enhanced sensitivity and imaging capability will enable the recovery of faint features — such as quiet Sun free–free emission, faint gyrosynchrotron emission from CMEs, and the narrow-band faint emissions, known as Weak Impulsive Narrowband Quiet Sun Emission \citep[WINQSEs;][]{Mondal_2020}  — even in the presence of bright active regions. For more information in this regard, refer to the chapter by \citet{Mondal01.2026.SKA}.
Recent work by \citet{Dey2025linear_pol} has demonstrated that although Stokes I sources may appear morphologically simple, the corresponding Stokes Q, U, and V maps reveal significant spatial, temporal, and spectral structure. Such behaviours will be captured even more effectively with SKA, owing to its high temporal and angular resolution and its ability to achieve spectral resolutions down to a few hertz in zoom modes.
The SKA-Low array (operating within 50–350 MHz) will provide a contiguous 300-MHz instantaneous bandwidth, while SKA-Mid will cover 0.35–15.4 GHz across four operational bands: Band 1 (350–1050 MHz), Band 2 (950–1760 MHz), Band 5a (4.6–8.5 GHz), and Band 5b (8.3–15.4 GHz). SKA telescopes can be configured into up to 16 independent subarrays, enabling simultaneous observations across a wide frequency range. Figure \ref{fig:band_coverage} illustrates the frequency-band coverage of existing radio interferometers for which polarimetric observations have been reported. Figure \ref{fig:time_freq_resolution} reveals the general trade-off between frequency and time resolution across instruments. While the above plots present the capabilities of currently operational instruments, earlier facilities such as Culgoora \citep{culgoora_1967}, Clark Lake \citep{1983_Clark-Lake}, and NoRH \citep{NoRH_Nakajima_1994} — now decommissioned — were also capable of carrying out spectro-polarimetric imaging studies of the corona.

As discussed in Section \ref{sec:mag_field_measurement}, studies of gyroresonance and free–free emission require wide frequency coverage for accurate magnetic-field diagnostics. Similarly, plasma emissions -- typically spanning broad metric to decimetric wavelength ranges -- demand a wide instantaneous bandwidth. With the SKA, simultaneous observations of solar emission across a wide range of coronal heights and into the inner heliosphere will become possible, offering transformative opportunities for coronal magnetic field measurements.

\begin{figure}[!h]
    \centering
    \includegraphics[width=\linewidth]{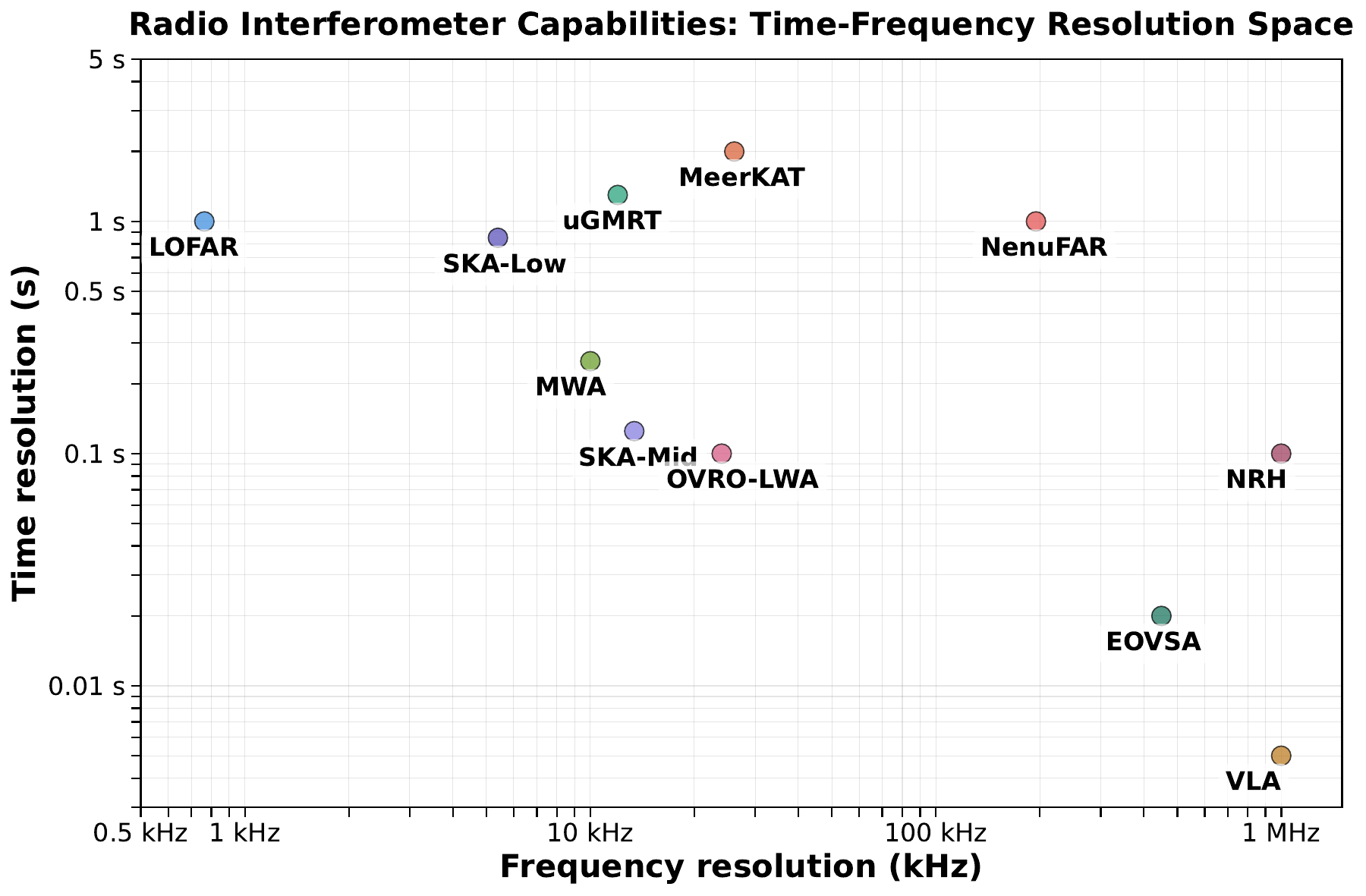}
    \caption{ The observational characteristics of radio interferometers, currently in operation for solar observation, across different frequency and time resolutions. The horizontal axis (logarithmic scale) represents frequency resolution, ranging from 0.5 kHz to 1 MHz, where higher values indicate coarser frequency resolution. The vertical axis (logarithmic scale) shows time resolution, ranging from 0.003 s to 5 s, with higher values indicating coarser temporal resolution.}
    \label{fig:time_freq_resolution}
\end{figure}

\begin{figure}[!h]
    \centering
    \includegraphics[width=0.9\linewidth]{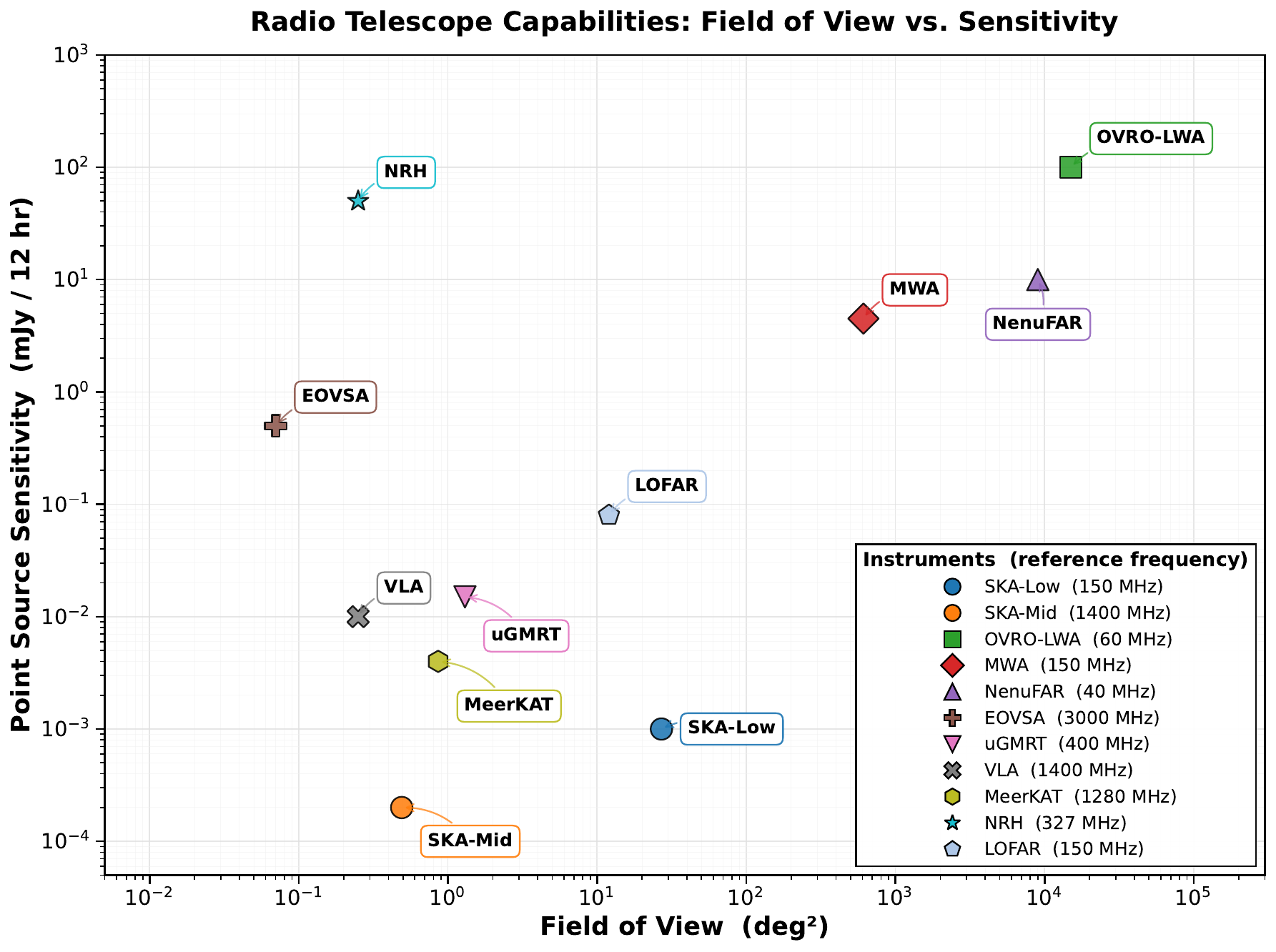}
    \caption{Comparison of radio interferometers, currently in operation, in the Field of View (FOV) vs. Point Source Sensitivity plane. Each instrument is represented by a unique marker and color, plotted according to its instantaneous FOV (x-axis, deg²) and point-source sensitivity achieved in 12 hours of integration (y-axis, mJy) at representative centre frequencies. The instruments span frequencies from $\sim40$ MHz (NenuFAR) to $\sim 3$ GHz (EOVSA), exhibiting dramatically different sensitivities and FOVs based on their physical design: aperture arrays (SKA-Low, MWA, NenuFAR, OVRO-LWA) achieve large FOVs but modest sensitivity, while traditional dish arrays (SKA-Mid, VLA, MeerKAT) provide higher sensitivity over smaller fields. Point-source sensitivities are estimated using $\sigma = \frac{\text{SEFD}}{\eta \sqrt{2 \Delta \nu t_{\text{int}}}} \quad \text{with} \quad \eta = 0.9$ and representative SEFD values from published instrument specifications. This figure is styled after Figure 12 from \citet{Tingay2013_MWA}, adapted to include current facility specifications.}
    \label{fig:time_freq_resolution}
\end{figure}
\subsection{Co-ordinated Radio and EUV/soft X-ray observations}
Several attempts have been made to measure the coronal magnetic fields with coordinated observations from radio and space-based instruments. \citet{brosius2002measurements} had measured the 3D coronal magnetic fields ($B(x,y,T)$) with the Very Large Array and with the Coronal Diagnostic Spectrometer, the EUV Imaging Telescope, and the Michelson Doppler Imager aboard the NASA/ESA Solar and Heliospheric Observatory satellite. They used the temperature as a proxy for the height in the corona. \citet{chen2020measurement} used MK4 coronameter data to constrain the electron density, while \citet{Kumari2019} used space-born whitelight coronagraph images to constrain the electron density for the estimation of magnetic field strength following this approach. Similarly, \citet{Gopalswamy2012} have used EUV images from SDO/AIA to measure shock standoff distances and derive Alfv\'enic Mach numbers and thus estimated the magnetic field.
With wideband spectro-polarimetric images available from the SKAO, coordinated observations with space-based instruments have a huge potential for 3D magnetic field modelling of the global corona as well as active regions.
Coordinated observations between SKAO and modern heliophysics missions, such as Solar Orbiter \citep{muller2020}, PUNCH \citep{DeForest2025_punch}, PROBA3 \citep{Zhukov2025ASPIICS}, Aditya-L1 \citep{Tripathi2023AdityaL1}, and PSP, offer a multi-wavelength pathway towards coronal diagnostics. When paired with the detailed temperature, density, and elemental abundance information from Solar Orbiter’s EUV/X-ray spectrometers, SKA-Mid can provide fully constrained 2D and even 3D coronal magnetograms across active regions and the quiet Sun. Meanwhile, PUNCH’s continuous wide-field imaging of the outer corona complements SKAO by tracking CMEs, shocks, and density structures as they propagate outward, allowing magnetic and plasma parameters to be connected directly to their large-scale evolutionary signatures. Instruments like Aditya L1, Solar Orbiter, PUNCH are expected to be operational when SKA becomes operational. Alongside these, a few upcoming missions will increase the capabilities of coordinated observation. For example: 
\begin{enumerate}
    \item NASA's MUSE \citep[Multi-slit Solar Explorer,][]{de2022probing} will deliver high-cadence EUV spectroscopic measurements of coronal temperature, density, and flows, providing essential plasma diagnostics for magnetic field estimation.
    \item ESA's Vigil \citep{palomba2022vigil} will observe the Sun from the L5 vantage point, providing complementary perspectives for stereoscopic reconstruction of active-region and CME magnetic topologies.
    \item NASA's HelioSwarm \citep{klein2023helioswarm} will characterize multi-scale solar-wind turbulence and magnetic fluctuations, improving our understanding of radio-wave propagation effects and the evolution of heliospheric magnetic fields.
\end{enumerate}
 Together, these facilities form a multi-layered diagnostic system, connecting magnetic and thermal conditions at the Sun to their evolution in the extended corona and inner heliosphere. This coordinated approach provides constraints needed to probe the magnetic drivers of solar variability and space weather with unprecedented completeness.

\section{Summary}\label{sec:summary}

Coronal magnetic field measurements remain one of the most critical challenges in solar and space-weather research. While traditional photospheric magnetography is well-established through Zeeman effect measurements, coronal field measurements face fundamental obstacles: extremely weak Zeeman splittings, optically thin emission, and line-of-sight cancellation effects that render optical techniques largely ineffective above the photosphere.

Radio observations provide unique diagnostic capabilities across multiple emission mechanisms and coronal heights. Free-free emission acquires weak circular polarization in magnetized plasma, enabling line-of-sight magnetic field measurements through the refractive index difference between extraordinary and ordinary modes. Gyroresonance emission at centimetric wavelengths directly probes strong magnetic fields (10²-10³ G) in narrow iso-Gauss layers above active regions, while gyrosynchrotron radiation from mildly relativistic electrons in flaring loops and CMEs provides diagnostics of fields ranging from sub-Gauss to hundreds of Gauss at coronal heights. Plasma emission mechanisms—manifested in type II, III, and IV radio bursts—offer both direct (through polarization signatures) and indirect (through shock dynamics and band-splitting) constraints on magnetic field strength and topology from the low corona into the heliosphere. Propagation effects, particularly mode coupling in quasi-transverse regions, provide additional diagnostic pathways. The circular polarization inversion observed when electromagnetic waves traverse transverse magnetic fields enables estimation of field strength at different coronal heights through spectral analysis of the critical depolarization frequency.

The SKA telescopes will transform coronal magnetography through unprecedented capabilities: (1) wide instantaneous bandwidth coverage (50-350 MHz for SKA-Low; 0.35-15.4 GHz for SKA-Mid) enabling simultaneous multi-height observations, (2) high sensitivity recovering faint emissions alongside bright active regions, (3) excellent spectro-temporal resolution capturing rapid magnetic evolution, and (4) superior imaging fidelity through dense uv coverage. Combined with coordinated observations from space-based missions (Solar Orbiter, PUNCH, PROBA-3, Aditya-L1, Parker Solar Probe), SKA will enable comprehensive 2D and 3D coronal magnetic field mapping across quiet Sun, active regions, and eruptive events—bridging the critical gap between photospheric measurements and heliospheric observations for understanding solar variability and space weather drivers.
\section{Acknowledgment}
This work presents observations from the MWA from Inyarrimanha Ilgari Bundara, the CSIRO Murchison Radio-astronomy Observatory. We acknowledge the Wajarri Yamaji people as the traditional owners and native title holders of the observatory site. This work also presents observations from MeerKAT radio telescope operated by the South African Radio Astronomy Observatory, which is a facility of the National Research Foundation, an agency of the Department of Science, Technology, and Innovation. This work shows data from the International LOFAR Telescope (ILT). LOFAR is designed and constructed by ASTRON. D.P., P.M., S.D., D.O. and S.M. acknowledge support from the Department of Atomic Energy, Government of India, under project 12-R\&D-TFR-5.02-0700. D.K. acknowledges financial support from the Severo Ochoa grant CEX2021-001131-S and from the Spanish grant PID2023-147883NB-C21, funded by MCIU/AEI/ 10.13039/501100011033, as well as support through ERDF/EU, and the NASA Living with a Star Jack Eddy Postdoctoral Fellowship Program, administered by UCAR’s Cooperative Programs for the Advancement of Earth System Science (CPAESS) under award 80NSSC22M0097.
\bibliographystyle{abbrvnat-maxbibnames4}
\bibliography{chapter}

\begin{thebibliography}{171}
\providecommand{\natexlab}[1]{#1}
\providecommand{\url}[1]{\texttt{#1}}
\expandafter\ifx\csname urlstyle\endcsname\relax
  \providecommand{\doi}[1]{doi: #1}\else
  \providecommand{\doi}{doi: \begingroup \urlstyle{rm}\Url}\fi

\bibitem[{Akhmedov} et~al.(1982){Akhmedov}, {Gelfreikh}, {Bogod}, and {Korzhavin}]{akhmedov_1982}
S.~B. {Akhmedov}, G.~B. {Gelfreikh}, V.~M. {Bogod}, and A.~N. {Korzhavin}.
\newblock \emph{\solphys}, 79\penalty0 (1):\penalty0 41--58, July 1982.
\newblock \doi{10.1007/BF00146972}.

\bibitem[Alissandrakis et~al.(1980)Alissandrakis, Kundu, and Lantos]{alissandrakis1980model}
C.~Alissandrakis, M.~Kundu, and P.~Lantos.
\newblock \emph{Astronomy and Astrophysics, vol. 82, no. 1-2, Feb. 1980, p. 30-40.}, 82:\penalty0 30--40, 1980.

\bibitem[Alissandrakis et~al.(1996)Alissandrakis, Borgioli, Chiuderi~Drago, Hagyard, and Shibasaki]{alissandrakis1996coronal}
C.~Alissandrakis et al.
\newblock \emph{Solar Physics}, 167\penalty0 (1):\penalty0 167--179, 1996.

\bibitem[Alissandrakis et~al.(2019)Alissandrakis, Bogod, Kaltman, Patsourakos, and Peterova]{alissandrakis2019modeling}
C.~Alissandrakis et al.
\newblock \emph{Solar Physics}, 294\penalty0 (2):\penalty0 23, 2019.

\bibitem[Alissandrakis and Gary(2021)]{alissandrakis2021radio}
C.~E. Alissandrakis and D.~E. Gary.
\newblock \emph{Frontiers in Astronomy and Space Sciences}, 7:\penalty0 591075, 2021.

\bibitem[{Alissandrakis} et~al.(2021){Alissandrakis}, {Nindos}, {Patsourakos}, and {Hillaris}]{Alissandrakis2021}
C.~E. {Alissandrakis}, A.~{Nindos}, S.~{Patsourakos}, and A.~{Hillaris}.
\newblock \emph{\aap}, 654:\penalty0 A112, Oct. 2021.
\newblock \doi{10.1051/0004-6361/202141672}.

\bibitem[Aurass et~al.(2005)Aurass, Rausche, Mann, and Hofmann]{aurass2005fiber}
H.~Aurass, G.~Rausche, G.~Mann, and A.~Hofmann.
\newblock \emph{Astronomy \& Astrophysics}, 435\penalty0 (3):\penalty0 1137--1148, 2005.

\bibitem[{Bain} et~al.(2014){Bain}, {Krucker}, {Saint-Hilaire}, and {Raftery}]{Bain2014}
H.~M. {Bain}, S.~{Krucker}, P.~{Saint-Hilaire}, and C.~L. {Raftery}.
\newblock \emph{\apj}, 782\penalty0 (1):\penalty0 43, February 2014.
\newblock \doi{10.1088/0004-637X/782/1/43}.

\bibitem[Bandiera(1982)]{bandiera1982diagnostic}
R.~Bandiera.
\newblock \emph{Astronomy and Astrophysics, vol. 112, no. 1, Aug. 1982, p. 52-60.}, 112:\penalty0 52--60, 1982.

\bibitem[{Bastian} et~al.(2001){Bastian}, {Pick}, {Kerdraon}, {Maia}, and {Vourlidas}]{bastian2001}
T.~S. {Bastian} et al.
\newblock \emph{\apjl}, 558\penalty0 (1):\penalty0 L65--L69, September 2001.
\newblock \doi{10.1086/323421}.

\bibitem[Benz and Mann(1998)]{benz1998intermediate}
A.~O. Benz and G.~Mann.
\newblock \emph{Astronomy and Astrophysics, v. 333, p. 1034-1042 (1998)}, 333:\penalty0 1034--1042, 1998.

\bibitem[{Benz} and {Zlobec}(1978)]{BenzZlobec1978}
A.~O. {Benz} and P.~{Zlobec}.
\newblock \emph{\aap}, 63\penalty0 (1-2):\penalty0 137--145, Feb. 1978.

\bibitem[{Benz} et~al.(1982){Benz}, {Treumann}, {Vilmer}, {Mangeney}, {Pick}, and {Raoult}]{Benz1982}
A.~O. {Benz} et al.
\newblock \emph{\aap}, 108\penalty0 (1):\penalty0 161--168, Apr. 1982.

\bibitem[Bernold and Treumann(1983)]{bernold1983fiber}
T.~E. Bernold and R.~A. Treumann.
\newblock \emph{Astrophysical Journal, Part 1, vol. 264, Jan. 15, 1983, p. 677-688. Research supported by the Swiss National Science Foundation and Swiss Society for Astronomy and Astrophysics.}, 264:\penalty0 677--688, 1983.

\bibitem[{Bhunia} et~al.(2023){Bhunia}, {Carley}, {Oberoi}, and {Gallagher}]{Bhunia2023}
S.~{Bhunia}, E.~P. {Carley}, D.~{Oberoi}, and P.~T. {Gallagher}.
\newblock \emph{\aap}, 670:\penalty0 A169, Feb. 2023.
\newblock \doi{10.1051/0004-6361/202244456}.

\bibitem[Bogod and Gelfreikh(1980)]{bogod1980measurements}
V.~Bogod and G.~Gelfreikh.
\newblock \emph{Solar Physics}, 67\penalty0 (1):\penalty0 29--46, 1980.

\bibitem[{Boischot}(1957)]{Boischot1957}
A.~{Boischot}.
\newblock \emph{Academie des Sciences Paris Comptes Rendus}, 244:\penalty0 1326--1329, January 1957.

\bibitem[{Brchnelova} et~al.(2023){Brchnelova}, {Kuźma}, {Zhang}, {Lani}, and {Poedts}]{Brchnelova23}
M.~{Brchnelova} et al.
\newblock \emph{A\&A}, 676:\penalty0 A83, 2023.
\newblock \doi{10.1051/0004-6361/202346788}.
\newblock URL \url{https://doi.org/10.1051/0004-6361/202346788}.

\bibitem[Brosius et~al.(2002)Brosius, Landi, Cook, Newmark, Gopalswamy, and Lara]{brosius2002measurements}
J.~W. Brosius et al.
\newblock \emph{The Astrophysical Journal}, 574\penalty0 (1):\penalty0 453, 2002.

\bibitem[{Carley} et~al.(2017){Carley}, {Vilmer}, {Sim{\~o}es}, and {{\'O} Fearraigh}]{Carley2017}
E.~P. {Carley}, N.~{Vilmer}, P.~J.~A. {Sim{\~o}es}, and B.~{{\'O} Fearraigh}.
\newblock \emph{\aap}, 608:\penalty0 A137, December 2017.
\newblock \doi{10.1051/0004-6361/201731368}.

\bibitem[Chen et~al.(2020)Chen, Shen, Gary, Reeves, Fleishman, Yu, Guo, Krucker, Lin, Nita, et~al.]{chen2020measurement}
B.~Chen et al.
\newblock \emph{Nature Astronomy}, 4\penalty0 (12):\penalty0 1140--1147, 2020.

\bibitem[Chiuderi~Drago et~al.(1987)Chiuderi~Drago, Alissandrakis, and Hagyard]{chiuderi1987microwave}
F.~Chiuderi~Drago, C.~Alissandrakis, and M.~Hagyard.
\newblock \emph{Solar physics}, 112\penalty0 (1):\penalty0 89--105, 1987.

\bibitem[{Cho} et~al.(2011){Cho}, {Bong}, {Moon}, {Shanmugaraju}, {Kwon}, and {Park}]{Cho2011}
K.-S. {Cho} et al.
\newblock \emph{\aap}, 530:\penalty0 A16, June 2011.
\newblock \doi{10.1051/0004-6361/201015578}.

\bibitem[Cohen(1960)]{cohen1960magnetoionic}
M.~H. Cohen.
\newblock \emph{Astrophysical Journal, vol. 131, p. 664}, 131:\penalty0 664, 1960.

\bibitem[Daei et~al.(2023)Daei, Pomoell, Price, Kumari, Good, and Kilpua]{daei2023modeling}
F.~Daei et al.
\newblock \emph{Astronomy \& Astrophysics}, 676:\penalty0 A141, 2023.

\bibitem[De~Pontieu et~al.(2022)De~Pontieu, Testa, Martinez-Sykora, Antolin, Karampelas, Hansteen, Rempel, Cheung, Reale, Danilovic, et~al.]{de2022probing}
B.~De~Pontieu et al.
\newblock \emph{The Astrophysical Journal}, 926\penalty0 (1):\penalty0 52, 2022.

\bibitem[{de Villiers}(2023)]{MeerKAT_primary_beam_Villier_2023}
M.~S. {de Villiers}.
\newblock \emph{\aj}, 165\penalty0 (3):\penalty0 78, Mar. 2023.
\newblock \doi{10.3847/1538-3881/acabc3}.

\bibitem[DeForest et~al.(2025)DeForest, Gibson, Killough, Waltham, Beasley, Colaninno, Laurent, Seaton, Hughes, Guhathakurta, Viall, Attie, Banerjee, Barnar, Biesecker, Bisi, Bothmer, Brody, Burkepile, Cairns, Campbell, Cheney, Case, Caspi, Chhiber, Clapp, Cranmer, Davies, de~Koning, Desai, Elliott, Farid, Gallardo-Lacourt, Gilly, Gobat, Hanson, Harrison, Hassler, Henley, Henry, Howard, Jackson, Jones, Kolinski, Lamb, Lehtinen, Lowder, Malanushenko, Matthaeus, McComas, McGee, Morgan, Oberoi, Odstrcil, Parmenter, Patel, Pecora, Persyn, Pizzo, Plunkett, Provornikova, Raouafi, Redfern, Rouillard, Smith, Smith, Talpas, Tappin, Thernisien, Thompson, Van~Kooten, Walsh, Webb, Wells, West, Wiens, and Yang]{DeForest2025_punch}
C.~DeForest et al.
\newblock \emph{arXiv e-prints}, page arXiv:2509.15131, 2025.
\newblock \doi{10.48550/arXiv.2509.15131}.

\bibitem[{Dey} et~al.(2025){Dey}, {Kansabanik}, {Oberoi}, and {Mondal}]{Dey2025linear_pol}
S.~{Dey}, D.~{Kansabanik}, D.~{Oberoi}, and S.~{Mondal}.
\newblock \emph{\apjl}, 988\penalty0 (2):\penalty0 L73, Aug. 2025.
\newblock \doi{10.3847/2041-8213/adef0e}.

\bibitem[{Dey, S.} et~al.(2025){Dey, S.}, {Oberoi, D.}, {Zucca, P.}, {Mancini, M.}, {Patra, D.}, and {Kansabanik, D.}]{Dey2025SIMPL}
{Dey, S.} et al.
\newblock \emph{Astronomy \& Astrophysics}, 704:\penalty0 A75, 2025.
\newblock \doi{10.1051/0004-6361/202556857}.
\newblock URL \url{https://doi.org/10.1051/0004-6361/202556857}.

\bibitem[Downs et~al.(2025)Downs, Linker, Caplan, Mason, Riley, Davidson, Reyes, Palmerio, Lionello, Turtle, et~al.]{downs2025near}
C.~Downs et al.
\newblock \emph{Science}, page eadq0872, 2025.

\bibitem[{Du} et~al.(2014){Du}, {Chen}, {Lv}, {Kong}, {Feng}, {Guo}, and {Li}]{Du2014}
G.~{Du} et al.
\newblock \emph{\apjl}, 793\penalty0 (2):\penalty0 L39, Oct. 2014.
\newblock \doi{10.1088/2041-8205/793/2/L39}.

\bibitem[{Dulk}(1985)]{dulk_ff}
G.~A. {Dulk}.
\newblock \emph{\araa}, 23:\penalty0 169--224, Jan. 1985.
\newblock \doi{10.1146/annurev.aa.23.090185.001125}.

\bibitem[{Dulk} and {Suzuki}(1980)]{DulkSuzuki1980}
G.~A. {Dulk} and S.~{Suzuki}.
\newblock \emph{\aap}, 88\penalty0 (1-2):\penalty0 203--217, Aug. 1980.

\bibitem[{Dulk} et~al.(1976){Dulk}, {Jacques}, {Smerd}, {MacQueen}, {Gosling}, {Steward}, {Sheridan}, {Robinson}, and {Magun}]{Dulk1976}
G.~A. {Dulk} et al.
\newblock \emph{\solphys}, 49:\penalty0 369, Aug. 1976.

\bibitem[{Duncan}(1981)]{Duncan1981}
R.~A. {Duncan}.
\newblock \emph{\solphys}, 73\penalty0 (1):\penalty0 191--204, Sept. 1981.
\newblock \doi{10.1007/BF00153154}.

\bibitem[{Fleishman} and {Melnikov}(2003)]{fleishmann_2003}
G.~D. {Fleishman} and V.~F. {Melnikov}.
\newblock \emph{\apj}, 587\penalty0 (2):\penalty0 823--835, Apr. 2003.
\newblock \doi{10.1086/368252}.

\bibitem[Fleishman et~al.(2020)Fleishman, Gary, Chen, Kuroda, Yu, and Nita]{fleishman2020decay}
G.~D. Fleishman et al.
\newblock \emph{Science}, 367\penalty0 (6475):\penalty0 278--280, 2020.

\bibitem[{Fleishman} et~al.(2021){Fleishman}, {Kuznetsov}, and {Landi}]{fleishman_ff}
G.~D. {Fleishman}, A.~A. {Kuznetsov}, and E.~{Landi}.
\newblock \emph{\apj}, 914\penalty0 (1):\penalty0 52, June 2021.
\newblock \doi{10.3847/1538-4357/abf92c}.

\bibitem[Gary and Hurford(1994)]{gary1994coronal}
D.~E. Gary and G.~Hurford.
\newblock \emph{The Astrophysical Journal, Part 1 (ISSN 0004-637X), vol. 420, no. 2, p. 903-912}, 420:\penalty0 903--912, 1994.

\bibitem[{Gary} et~al.(2018){Gary}, {Chen}, {Dennis}, {Fleishman}, {Hurford}, {Krucker}, {McTiernan}, {Nita}, {Shih}, {White}, and {Yu}]{Gary2018_EOVSA}
D.~E. {Gary} et al.
\newblock \emph{\apj}, 863\penalty0 (1):\penalty0 83, Aug. 2018.
\newblock \doi{10.3847/1538-4357/aad0ef}.

\bibitem[Gary et~al.(2018)Gary, Chen, Dennis, Fleishman, Hurford, Krucker, McTiernan, Nita, Shih, White, et~al.]{gary2018microwave}
D.~E. Gary et al.
\newblock \emph{The Astrophysical Journal}, 863\penalty0 (1):\penalty0 83, 2018.

\bibitem[Gary(2001)]{gary2001plasma}
G.~A. Gary.
\newblock \emph{Solar Physics}, 203\penalty0 (1):\penalty0 71--86, 2001.

\bibitem[Gelfreikh et~al.(1987)Gelfreikh, Peterova, and Ryabov]{gelfreikh1987measurements}
G.~Gelfreikh, N.~Peterova, and B.~Ryabov.
\newblock \emph{Solar physics}, 108\penalty0 (1):\penalty0 89--97, 1987.

\bibitem[Gibson et~al.(2016)Gibson, Kucera, White, Dove, Fan, Forland, Rachmeler, Downs, and Reeves]{Gibson2016}
S.~E. Gibson et al.
\newblock \emph{Frontiers in Astronomy and Space Sciences}, Volume 3 - 2016, 2016.
\newblock ISSN 2296-987X.
\newblock \doi{10.3389/fspas.2016.00008}.
\newblock URL \url{https://www.frontiersin.org/journals/astronomy-and-space-sciences/articles/10.3389/fspas.2016.00008}.

\bibitem[Ginzburg et~al.(1962)Ginzburg, Sadowski, Gallik, and Brown]{ginzburg1962propagation}
V.~L. Ginzburg, W.~L. Sadowski, D.~Gallik, and S.~C. Brown.
\newblock \emph{Physics Today}, 15\penalty0 (10):\penalty0 70--73, 1962.

\bibitem[{Gopalswamy} and {Kundu}(1987)]{Gopalswamy_kundu1987}
N.~{Gopalswamy} and M.~R. {Kundu}.
\newblock \emph{\solphys}, 114\penalty0 (2):\penalty0 347--362, Sept. 1987.
\newblock \doi{10.1007/BF00167350}.

\bibitem[{Gopalswamy} et~al.(2012){Gopalswamy}, {Nitta}, {Akiyama}, {M{\"a}kel{\"a}}, and {Yashiro}]{Gopalswamy2012}
N.~{Gopalswamy} et al.
\newblock \emph{\apj}, 744\penalty0 (1):\penalty0 72, January 2012.
\newblock \doi{10.1088/0004-637X/744/1/72}.

\bibitem[{Hallinan} et~al.(2023){Hallinan}, {Anderson}, {Isella}, {Gary}, {Bowman}, {Romero-Wolf}, and {OVRO-LWA Collaboration}]{ovro_lwa_Hallinan_2023}
G.~{Hallinan} et al.
\newblock In \emph{American Astronomical Society Meeting Abstracts \#241}, volume 241 of \emph{American Astronomical Society Meeting Abstracts}, page 451.09, Jan. 2023.

\bibitem[{Hamaker} et~al.(1996){Hamaker}, {Bregman}, and {Sault}]{Hamakar1996}
J.~P. {Hamaker}, J.~D. {Bregman}, and R.~J. {Sault}.
\newblock \emph{\aaps}, 117:\penalty0 137--147, May 1996.

\bibitem[{Hariharan} et~al.(2014){Hariharan}, {Ramesh}, {Kishore}, {Kathiravan}, and {Gopalswamy}]{Hariharan2014}
K.~{Hariharan} et al.
\newblock \emph{\apj}, 795\penalty0 (1):\penalty0 14, Nov. 2014.
\newblock \doi{10.1088/0004-637X/795/1/14}.

\bibitem[{Hariharan} et~al.(2015){Hariharan}, {Ramesh}, and {Kathiravan}]{Hariharan2015}
K.~{Hariharan}, R.~{Ramesh}, and C.~{Kathiravan}.
\newblock \emph{\solphys}, 290\penalty0 (9):\penalty0 2479--2489, Sept. 2015.
\newblock \doi{10.1007/s11207-015-0761-5}.

\bibitem[{Hariharan} et~al.(2016){Hariharan}, {Ramesh}, {Kathiravan}, and {Wang}]{hariharan2016}
K.~{Hariharan}, R.~{Ramesh}, C.~{Kathiravan}, and T.~J. {Wang}.
\newblock \emph{\solphys}, 291\penalty0 (5):\penalty0 1405--1416, May 2016.
\newblock \doi{10.1007/s11207-016-0918-x}.

\bibitem[{Hicks} et~al.(2012){Hicks}, {Paravastu-Dalal}, {Stewart}, {Erickson}, {Ray}, {Kassim}, {Burns}, {Clarke}, {Schmitt}, {Craig}, {Hartman}, and {Weiler}]{nenufar_Hicks2012}
B.~C. {Hicks} et al.
\newblock \emph{\pasp}, 124\penalty0 (920):\penalty0 1090, Oct. 2012.
\newblock \doi{10.1086/668121}.

\bibitem[Iwai et~al.(2014)Iwai, Shibasaki, Nozawa, Takahashi, Sawada, Kitagawa, Miyawaki, and Kashiwagi]{iwai2014coronal}
K.~Iwai et al.
\newblock \emph{Earth, Planets and Space}, 66\penalty0 (1):\penalty0 149, 2014.

\bibitem[Jonas and Team(2016)]{jonas2016meerkat}
J.~Jonas and M.~Team.
\newblock \emph{MeerKAT science: on the pathway to the SKA}, page~1, 2016.

\bibitem[{Kai}(1969)]{kai1969}
K.~{Kai}.
\newblock \emph{\pasa}, 1\penalty0 (5):\penalty0 189--191, Mar. 1969.
\newblock \doi{10.1017/S1323358000011383}.

\bibitem[Kakinuma and Swarup(1962)]{kakinuma1962model}
T.~Kakinuma and G.~Swarup.
\newblock \emph{Astrophysical Journal, vol. 136, p. 975}, 136:\penalty0 975, 1962.

\bibitem[Kansabanik(2022)]{kansabanik2022working}
D.~Kansabanik.
\newblock \emph{Solar Physics}, 297\penalty0 (9):\penalty0 122, 2022.

\bibitem[{Kansabanik} et~al.(2022{\natexlab{a}}){Kansabanik}, {Oberoi}, and {Mondal}]{paircars22a}
D.~{Kansabanik}, D.~{Oberoi}, and S.~{Mondal}.
\newblock \emph{\apj}, 932\penalty0 (2):\penalty0 110, June 2022{\natexlab{a}}.
\newblock \doi{10.3847/1538-4357/ac6758}.

\bibitem[{Kansabanik} et~al.(2022{\natexlab{b}}){Kansabanik}, {Oberoi}, and {Mondal}]{paircars22b}
D.~{Kansabanik}, D.~{Oberoi}, and S.~{Mondal}.
\newblock \emph{arXiv e-prints}, art. arXiv:2207.11924, July 2022{\natexlab{b}}.
\newblock \doi{10.48550/arXiv.2207.11924}.

\bibitem[{Kansabanik} et~al.(2023{\natexlab{a}}){Kansabanik}, {Bera}, {Oberoi}, and {Mondal}]{paircars23}
D.~{Kansabanik}, A.~{Bera}, D.~{Oberoi}, and S.~{Mondal}.
\newblock \emph{\apjs}, 264\penalty0 (2):\penalty0 47, Feb. 2023{\natexlab{a}}.
\newblock \doi{10.3847/1538-4365/acac79}.

\bibitem[{Kansabanik} et~al.(2023{\natexlab{b}}){Kansabanik}, {Mondal}, and {Oberoi}]{Kansabanik2023_CME1}
D.~{Kansabanik}, S.~{Mondal}, and D.~{Oberoi}.
\newblock \emph{\apj}, 950\penalty0 (2):\penalty0 164, June 2023{\natexlab{b}}.
\newblock \doi{10.3847/1538-4357/acc385}.

\bibitem[{Kansabanik} et~al.(2024){Kansabanik}, {Mondal}, and {Oberoi}]{Kansabanik2024_gs_StokesV}
D.~{Kansabanik}, S.~{Mondal}, and D.~{Oberoi}.
\newblock \emph{\apj}, 968\penalty0 (2):\penalty0 55, June 2024.
\newblock \doi{10.3847/1538-4357/ad43e9}.

\bibitem[Kansabanik et~al.(2024)Kansabanik, Mondal, Oberoi, Chibueze, Engelbrecht, Strauss, Kontar, Botha, Steyn, and Nel]{Kansabanik_2024_meerKAT}
D.~Kansabanik et al.
\newblock \emph{The Astrophysical Journal}, 961\penalty0 (1):\penalty0 96, jan 2024.
\newblock \doi{10.3847/1538-4357/ad0b7f}.
\newblock URL \url{https://doi.org/10.3847/1538-4357/ad0b7f}.

\bibitem[Kansabanik et~al.(2025)Kansabanik, Gouws, Patra, Vourlidas, Kotzé, Oberoi, Shaik, Buchner, and Camilo]{kansabanik_2025_meerKAT}
D.~Kansabanik et al.
\newblock \emph{Frontiers in Astronomy and Space Sciences}, Volume 12 - 2025, 2025.
\newblock ISSN 2296-987X.
\newblock \doi{10.3389/fspas.2025.1666743}.
\newblock URL \url{https://www.frontiersin.org/journals/astronomy-and-space-sciences/articles/10.3389/fspas.2025.1666743}.

\bibitem[{Kansabanik} et~al.(2025){Kansabanik}, {Vourlidas}, {Dey}, {Mondal}, and {Oberoi}]{cross_phase_Kansabanik2025}
D.~{Kansabanik} et al.
\newblock \emph{\apjs}, 278\penalty0 (1):\penalty0 26, May 2025.
\newblock \doi{10.3847/1538-4365/adc443}.

\bibitem[Klein et~al.(2023)Klein, Spence, Alexandrova, Argall, Arzamasskiy, Bookbinder, Broeren, Caprioli, Case, Chandran, et~al.]{klein2023helioswarm}
K.~G. Klein et al.
\newblock \emph{Space Science Reviews}, 219\penalty0 (8):\penalty0 74, 2023.

\bibitem[{Klein} and {Trottet}(1984)]{Klein_Trottet1984}
K.-L. {Klein} and G.~{Trottet}.
\newblock \emph{\aap}, 141\penalty0 (1):\penalty0 67--76, Dec. 1984.

\bibitem[{Komesaroff}(1958)]{Komesaroff1958}
M.~{Komesaroff}.
\newblock \emph{Australian Journal of Physics}, 11:\penalty0 201, June 1958.
\newblock \doi{10.1071/PH580201}.

\bibitem[{Kouloumvakos} et~al.(2021){Kouloumvakos}, {Rouillard}, {Warmuth}, {Magdalenic}, {Jebaraj}, {Mann}, {Vainio}, and {Monstein}]{Kouloumvakos2021}
A.~{Kouloumvakos} et al.
\newblock \emph{\apj}, 913\penalty0 (2):\penalty0 99, June 2021.
\newblock \doi{10.3847/1538-4357/abf435}.

\bibitem[Kramers(1923)]{kramers1923xciii}
H.~A. Kramers.
\newblock \emph{The London, Edinburgh, and Dublin Philosophical Magazine and Journal of Science}, 46\penalty0 (275):\penalty0 836--871, 1923.

\bibitem[Kuijpers(1975)]{kuijpers1975generation}
J.~Kuijpers.
\newblock \emph{Solar Physics}, 44\penalty0 (1):\penalty0 173--193, 1975.

\bibitem[{Kumari} et~al.(2017{\natexlab{a}}){Kumari}, {Ramesh}, {Kathiravan}, and {Gopalswamy}]{Kumari2017a}
A.~{Kumari}, R.~{Ramesh}, C.~{Kathiravan}, and N.~{Gopalswamy}.
\newblock \emph{\apj}, 843\penalty0 (1):\penalty0 10, July 2017{\natexlab{a}}.
\newblock \doi{10.3847/1538-4357/aa72e7}.

\bibitem[{Kumari} et~al.(2017{\natexlab{b}}){Kumari}, {Ramesh}, {Kathiravan}, and {Wang}]{Kumari2017typeII_band}
A.~{Kumari}, R.~{Ramesh}, C.~{Kathiravan}, and T.~J. {Wang}.
\newblock \emph{\solphys}, 292\penalty0 (11):\penalty0 161, November 2017{\natexlab{b}}.
\newblock \doi{10.1007/s11207-017-1180-6}.

\bibitem[{Kumari} et~al.(2019){Kumari}, {Ramesh}, {Kathiravan}, {Wang}, and {Gopalswamy}]{Kumari2019}
A.~{Kumari} et al.
\newblock \emph{\apj}, 881\penalty0 (1):\penalty0 24, August 2019.
\newblock \doi{10.3847/1538-4357/ab2adf}.

\bibitem[Kundu and Alissandrakis(1984)]{kundu1984structure}
M.~Kundu and C.~Alissandrakis.
\newblock \emph{Solar physics}, 94\penalty0 (2):\penalty0 249--283, 1984.

\bibitem[{Kundu}(1965)]{kundu1965}
M.~R. {Kundu}.
\newblock In C.~{de Jager}, editor, \emph{The Solar Spectrum}, volume~1 of \emph{Astrophysics and Space Science Library}, page 408, Jan. 1965.
\newblock \doi{10.1007/978-94-010-3587-3_21}.

\bibitem[Kundu(1965)]{kundu1965solar}
M.~R. Kundu.
\newblock \emph{New York: Interscience Publication}, 1965.

\bibitem[{Kundu} et~al.(1983){Kundu}, {Erickson}, {Gergely}, {Mahoney}, and {Turner}]{1983_Clark-Lake}
M.~R. {Kundu} et al.
\newblock \emph{\solphys}, 83\penalty0 (2):\penalty0 385--389, Mar. 1983.
\newblock \doi{10.1007/BF00148288}.

\bibitem[Kuznetsov and Kontar(2015)]{kuznetsov2015spatially}
A.~Kuznetsov and E.~Kontar.
\newblock \emph{Solar Physics}, 290\penalty0 (1):\penalty0 79--93, 2015.

\bibitem[{Lagg} et~al.(2017){Lagg}, {Lites}, {Harvey}, {Gosain}, and {Centeno}]{Lagg2017}
A.~{Lagg} et al.
\newblock \emph{\ssr}, 210\penalty0 (1-4):\penalty0 37--76, Sept. 2017.
\newblock \doi{10.1007/s11214-015-0219-y}.

\bibitem[Lang et~al.(1993)Lang, Willson, Kile, Lemen, Strong, Bogod, Gelfreikh, Ryabov, Hafizov, Abramov, et~al.]{lang1993magnetospheres}
K.~Lang et al.
\newblock \emph{Astrophysical Journal v. 419, p. 398}, 419:\penalty0 398, 1993.

\bibitem[Lee(2007)]{lee2007radio}
J.~Lee.
\newblock \emph{Space Science Reviews}, 133\penalty0 (1):\penalty0 73--102, 2007.

\bibitem[{Lemen} et~al.(2012){Lemen}, {Title}, {Akin}, {Boerner}, {Chou}, {Drake}, {Duncan}, {Edwards}, {Friedlaender}, {Heyman}, {Hurlburt}, {Katz}, {Kushner}, {Levay}, {Lindgren}, {Mathur}, {McFeaters}, {Mitchell}, {Rehse}, {Schrijver}, {Springer}, {Stern}, {Tarbell}, {Wuelser}, {Wolfson}, {Yanari}, {Bookbinder}, {Cheimets}, {Caldwell}, {Deluca}, {Gates}, {Golub}, {Park}, {Podgorski}, {Bush}, {Scherrer}, {Gummin}, {Smith}, {Auker}, {Jerram}, {Pool}, {Soufli}, {Windt}, {Beardsley}, {Clapp}, {Lang}, and {Waltham}]{AIA_Lemen2012}
J.~R. {Lemen} et al.
\newblock \emph{\solphys}, 275\penalty0 (1-2):\penalty0 17--40, Jan. 2012.
\newblock \doi{10.1007/s11207-011-9776-8}.

\bibitem[Lin et~al.(2004)Lin, Kuhn, and Coulter]{lin2004coronal}
H.~Lin, J.~Kuhn, and R.~Coulter.
\newblock \emph{The Astrophysical Journal}, 613\penalty0 (2):\penalty0 L177, 2004.

\bibitem[Loukitcheva(2020)]{loukitcheva2020measuring}
M.~Loukitcheva.
\newblock \emph{Frontiers in Astronomy and Space Sciences}, 7:\penalty0 45, 2020.

\bibitem[Loukitcheva et~al.(2017)Loukitcheva, White, Solanki, Fleishman, and Carlsson]{loukitcheva2017millimeter}
M.~Loukitcheva et al.
\newblock \emph{Astronomy \& Astrophysics}, 601:\penalty0 A43, 2017.

\bibitem[{Mahrous} et~al.(2018){Mahrous}, {Alielden}, {Vr{\v{s}}nak}, and {Youssef}]{Mahrous2018}
A.~{Mahrous}, K.~{Alielden}, B.~{Vr{\v{s}}nak}, and M.~{Youssef}.
\newblock \emph{Journal of Atmospheric and Solar-Terrestrial Physics}, 172:\penalty0 75--82, July 2018.
\newblock \doi{10.1016/j.jastp.2018.03.018}.

\bibitem[{Maia} et~al.(2007){Maia}, {Gama}, {Mercier}, {Pick}, {Kerdraon}, and {Karlick{\'y}}]{Maia2007}
D.~J.~F. {Maia} et al.
\newblock \emph{\apj}, 660\penalty0 (1):\penalty0 874--881, May 2007.
\newblock \doi{10.1086/508011}.

\bibitem[McCauley et~al.(2019)McCauley, Cairns, White, Mondal, Lenc, Morgan, and Oberoi]{mccauley2019low}
P.~I. McCauley et al.
\newblock \emph{Solar Physics}, 294\penalty0 (8):\penalty0 106, 2019.

\bibitem[{Melrose} and {Dulk}(1982)]{MelroseDulk1982}
D.~B. {Melrose} and G.~A. {Dulk}.
\newblock \emph{\apj}, 259:\penalty0 844--858, Aug. 1982.
\newblock \doi{10.1086/160219}.

\bibitem[{Melrose} and {Sy}(1972)]{MelroseSy1972}
D.~B. {Melrose} and W.~N. {Sy}.
\newblock \emph{Australian Journal of Physics}, 25:\penalty0 387, Aug. 1972.
\newblock \doi{10.1071/PH720387}.

\bibitem[{Melrose} et~al.(1978){Melrose}, {Dulk}, and {Smerd}]{MelroseDulkSmerd1978}
D.~B. {Melrose}, G.~A. {Dulk}, and S.~F. {Smerd}.
\newblock \emph{\aap}, 66\penalty0 (3):\penalty0 315--324, June 1978.

\bibitem[{Melrose} et~al.(1980){Melrose}, {Dulk}, and {Gary}]{Melrose1980}
D.~B. {Melrose}, G.~A. {Dulk}, and D.~E. {Gary}.
\newblock \emph{\pasa}, 4\penalty0 (1):\penalty0 50--53, Jan. 1980.
\newblock \doi{10.1017/S1323358000018762}.

\bibitem[{Melrose}(1977)]{Melrose1977}
D.~V. {Melrose}.
\newblock \emph{Izvestiia Vysshaia Uchebn. Zaved., Radiofizika}, 20:\penalty0 1369--1378, Jan. 1977.

\bibitem[{Mercier}(1990)]{Mercier1990}
C.~{Mercier}.
\newblock \emph{\solphys}, 130\penalty0 (1-2):\penalty0 119--129, Dec. 1990.
\newblock \doi{10.1007/BF00156783}.

\bibitem[{Mondal} et~al.(2020{\natexlab{a}}){Mondal}, {Oberoi}, and {Mohan}]{Mondal_2020}
S.~{Mondal}, D.~{Oberoi}, and A.~{Mohan}.
\newblock \emph{\apjl}, 895\penalty0 (2):\penalty0 L39, June 2020{\natexlab{a}}.
\newblock \doi{10.3847/2041-8213/ab8817}.

\bibitem[{Mondal} et~al.(2020{\natexlab{b}}){Mondal}, {Oberoi}, and {Vourlidas}]{Mondal2020a}
S.~{Mondal}, D.~{Oberoi}, and A.~{Vourlidas}.
\newblock \emph{\apj}, 893\penalty0 (1):\penalty0 28, April 2020{\natexlab{b}}.
\newblock \doi{10.3847/1538-4357/ab7fab}.

\bibitem[Mondal et~al.(2026)Mondal, author2, author3, author4, and author5]{Mondal01.2026.SKA}
S.~Mondal et al.
\newblock In \emph{Advancing Astrophysics with the SKA -- II (AASKAII)}. 2026.
\newblock arXiv search: Report number AASKAII/Mondal01.

\bibitem[{Morosan} et~al.(2019){Morosan}, {Kilpua}, {Carley}, and {Monstein}]{morosan2019}
D.~E. {Morosan}, E.~K.~J. {Kilpua}, E.~P. {Carley}, and C.~{Monstein}.
\newblock \emph{\aap}, 623:\penalty0 A63, March 2019.
\newblock \doi{10.1051/0004-6361/201834510}.

\bibitem[{M{\"u}ller} et~al.(2020){M{\"u}ller}, {St. Cyr}, {Zouganelis}, {Gilbert}, {Marsden}, {Nieves-Chinchilla}, {Antonucci}, {Auch{\`e}re}, {Berghmans}, {Horbury}, {Howard}, {Krucker}, {Maksimovic}, {Owen}, {Rochus}, {Rodriguez-Pacheco}, {Romoli}, {Solanki}, {Bruno}, {Carlsson}, {Fludra}, {Harra}, {Hassler}, {Livi}, {Louarn}, {Peter}, {Sch{\"u}hle}, {Teriaca}, {del Toro Iniesta}, {Wimmer-Schweingruber}, {Marsch}, {Velli}, {De Groof}, {Walsh}, and {Williams}]{muller2020}
D.~{M{\"u}ller} et al.
\newblock \emph{\aap}, 642:\penalty0 A1, Oct. 2020.
\newblock \doi{10.1051/0004-6361/202038467}.

\bibitem[Nagelis and Ryabov(1992)]{nagelis1992energetics}
J.~Nagelis and B.~Ryabov.
\newblock \emph{Kinematics and Physics of Celestial Bodies}, 8\penalty0 (6):\penalty0 28--32, 1992.

\bibitem[{Nakajima} et~al.(1985){Nakajima}, {Sekiguchi}, {Sawa}, {Kai}, and {Kawashima}]{Nakajima1985Nobeyama}
H.~{Nakajima} et al.
\newblock \emph{\pasj}, 37\penalty0 (1):\penalty0 163--170, Mar. 1985.
\newblock \doi{10.1093/pasj/37.1.163}.

\bibitem[{Nakajima} et~al.(1994){Nakajima}, {Nishio}, {Enome}, {Shibasaki}, {Takano}, {Hanaoka}, {Torii}, {Sekiguchi}, {Bushimata}, {Kawashima}, {Shinohara}, {Irimajiri}, {Koshiishi}, {Kosugi}, {Shiomi}, {Sawa}, and {Kai}]{NoRH_Nakajima_1994}
H.~{Nakajima} et al.
\newblock \emph{IEEE Proceedings}, 82\penalty0 (5):\penalty0 705--713, May 1994.
\newblock \doi{10.1109/5.284737}.

\bibitem[Nindos et~al.(2000)Nindos, White, Kundu, and Gary]{nindos2000observations}
A.~Nindos, S.~White, M.~Kundu, and D.~Gary.
\newblock \emph{The Astrophysical Journal}, 533\penalty0 (2):\penalty0 1053, 2000.

\bibitem[{Nita} et~al.(2015){Nita}, {Fleishman}, {Kuznetsov}, {Kontar}, and {Gary}]{nita_2015}
G.~M. {Nita} et al.
\newblock \emph{\apj}, 799\penalty0 (2):\penalty0 236, Feb. 2015.
\newblock \doi{10.1088/0004-637X/799/2/236}.

\bibitem[Nita et~al.(2018)Nita, Viall, Klimchuk, Loukitcheva, Gary, Kuznetsov, and Fleishman]{nita2018dressing}
G.~M. Nita et al.
\newblock \emph{The Astrophysical Journal}, 853\penalty0 (1):\penalty0 66, 2018.

\bibitem[Oberoi et~al.(2026)Oberoi, author2, author3, author4, and author5]{Oberoi01.2026.SKA}
D.~Oberoi et al.
\newblock In \emph{Advancing Astrophysics with the SKA -- II (AASKAII)}. 2026.
\newblock arXiv search: Report number AASKAII/Oberoi01.

\bibitem[Palomba and Luntama(2022)]{palomba2022vigil}
M.~Palomba and J.-P. Luntama.
\newblock \emph{44th COSPAR Scientific Assembly. Held 16-24 July}, 44:\penalty0 3544, 2022.

\bibitem[{Perley} et~al.(2011){Perley}, {Chandler}, {Butler}, and {Wrobel}]{Perley2011_vla}
R.~A. {Perley}, C.~J. {Chandler}, B.~J. {Butler}, and J.~M. {Wrobel}.
\newblock \emph{\apjl}, 739\penalty0 (1):\penalty0 L1, Sept. 2011.
\newblock \doi{10.1088/2041-8205/739/1/L1}.

\bibitem[Perri et~al.(2022)Perri, Leitner, Brchnelova, Baratashvili, Kuźma, Zhang, Lani, and Poedts]{Perri22}
B.~Perri et al.
\newblock \emph{\apj}, 936\penalty0 (1):\penalty0 19, aug 2022.
\newblock \doi{10.3847/1538-4357/ac7237}.
\newblock URL \url{https://dx.doi.org/10.3847/1538-4357/ac7237}.

\bibitem[Perri et~al.(2023)Perri, Kuźma, Brchnelova, Baratashvili, Zhang, Leitner, Lani, and Poedts]{Perri23}
B.~Perri et al.
\newblock \emph{\apj}, 943\penalty0 (2):\penalty0 124, feb 2023.
\newblock \doi{10.3847/1538-4357/ac9799}.
\newblock URL \url{https://dx.doi.org/10.3847/1538-4357/ac9799}.

\bibitem[Peterova and Akhmedov(1974)]{peterova1974influence}
N.~Peterova and S.~B. Akhmedov.
\newblock \emph{Soviet Astronomy, Vol. 17, p. 768}, 17:\penalty0 768, 1974.

\bibitem[Piddington and Minnett(1951)]{piddington1951solar}
J.~H. Piddington and H.~Minnett.
\newblock \emph{Australian Journal of Chemistry}, 4\penalty0 (2):\penalty0 131--157, 1951.

\bibitem[{Rahman} et~al.(2020){Rahman}, {Cairns}, and {McCauley}]{Rahman2020}
M.~M. {Rahman}, I.~H. {Cairns}, and P.~I. {McCauley}.
\newblock \emph{\solphys}, 295\penalty0 (3):\penalty0 51, Mar. 2020.
\newblock \doi{10.1007/s11207-020-01616-0}.

\bibitem[{Ramaty}(1969)]{ramaty_gs_1969}
R.~{Ramaty}.
\newblock \emph{\apj}, 158:\penalty0 753, Nov. 1969.
\newblock \doi{10.1086/150235}.

\bibitem[{Ramesh} and {Kathiravan}(2022)]{Ramesh_and_Kathiravan2022}
R.~{Ramesh} and C.~{Kathiravan}.
\newblock \emph{\apj}, 940\penalty0 (1):\penalty0 80, November 2022.
\newblock \doi{10.3847/1538-4357/ac9c65}.

\bibitem[Ramesh et~al.(2010)Ramesh, Kathiravan, and Sastry]{ramesh2010estimation}
R.~Ramesh, C.~Kathiravan, and C.~V. Sastry.
\newblock \emph{The Astrophysical Journal}, 711\penalty0 (2):\penalty0 1029, 2010.

\bibitem[{Ramesh} et~al.(2022){Ramesh}, {Kathiravan}, and {Chellasamy}]{Ramesh_2022}
R.~{Ramesh}, C.~{Kathiravan}, and E.~E. {Chellasamy}.
\newblock \emph{\apj}, 932\penalty0 (1):\penalty0 48, June 2022.
\newblock \doi{10.3847/1538-4357/ac6f05}.

\bibitem[{Ramesh} et~al.(2023){Ramesh}, {Kathiravan}, and {Kumari}]{Ramesh_2023}
R.~{Ramesh}, C.~{Kathiravan}, and A.~{Kumari}.
\newblock \emph{\apj}, 943\penalty0 (1):\penalty0 43, January 2023.
\newblock \doi{10.3847/1538-4357/acaea5}.

\bibitem[{Raouafi} et~al.(2023){Raouafi}, {Matteini}, {Squire}, {Badman}, {Velli}, {Klein}, {Chen}, {Matthaeus}, {Szabo}, {Linton}, {Allen}, {Szalay}, {Bruno}, {Decker}, {Akhavan-Tafti}, {Agapitov}, {Bale}, {Bandyopadhyay}, {Battams}, {Ber{\v{c}}i{\v{c}}}, {Bourouaine}, {Bowen}, {Cattell}, {Chandran}, {Chhiber}, {Cohen}, {D'Amicis}, {Giacalone}, {Hess}, {Howard}, {Horbury}, {Jagarlamudi}, {Joyce}, {Kasper}, {Kinnison}, {Laker}, {Liewer}, {Malaspina}, {Mann}, {McComas}, {Niembro-Hernandez}, {Nieves-Chinchilla}, {Panasenco}, {Pokorn{\'y}}, {Pusack}, {Pulupa}, {Perez}, {Riley}, {Rouillard}, {Shi}, {Stenborg}, {Tenerani}, {Verniero}, {Viall}, {Vourlidas}, {Wood}, {Woodham}, and {Woolley}]{Raoutfi2023}
N.~E. {Raouafi} et al.
\newblock \emph{\ssr}, 219\penalty0 (1):\penalty0 8, Feb. 2023.
\newblock \doi{10.1007/s11214-023-00952-4}.

\bibitem[Rimmele et~al.(2020)Rimmele, Warner, Keil, Goode, Kn{\"o}lker, Kuhn, Rosner, McMullin, Casini, Lin, et~al.]{rimmele2020daniel}
T.~R. Rimmele et al.
\newblock \emph{Solar Physics}, 295\penalty0 (12):\penalty0 1--49, 2020.

\bibitem[Ryabov et~al.(1999)Ryabov, Pilyeva, Alissandrakis, Shibasaki, Bogod, Garaimov, and Gelfreikh]{ryabov1999coronal}
B.~Ryabov et al.
\newblock \emph{Solar Physics}, 185\penalty0 (1):\penalty0 157--175, 1999.

\bibitem[Ryabov et~al.(2005)Ryabov, Maksimov, Lesovoi, Shibasaki, Nindos, and Pevtsov]{ryabov2005coronal}
B.~Ryabov et al.
\newblock \emph{Solar Physics}, 226\penalty0 (2):\penalty0 223--237, 2005.

\bibitem[{Salas-Matamoros} and {Klein}(2020)]{Salas_and_ludwig2020}
C.~{Salas-Matamoros} and K.-L. {Klein}.
\newblock \emph{\aap}, 639:\penalty0 A102, July 2020.
\newblock \doi{10.1051/0004-6361/202037989}.

\bibitem[{Sastry}(2009)]{sastry_circular}
C.~V. {Sastry}.
\newblock \emph{\apj}, 697\penalty0 (2):\penalty0 1934--1939, June 2009.
\newblock \doi{10.1088/0004-637X/697/2/1934}.

\bibitem[{Sault} et~al.(1996){Sault}, {Hamaker}, and {Bregman}]{Sault1996}
R.~J. {Sault}, J.~P. {Hamaker}, and J.~D. {Bregman}.
\newblock \emph{\aaps}, 117:\penalty0 149--159, May 1996.

\bibitem[Schad et~al.(2024)Schad, Petrie, Kuhn, Fehlmann, Rimmele, Tritschler, Woeger, Scholl, Williams, Harrington, et~al.]{schad2024mapping}
T.~A. Schad et al.
\newblock \emph{Science Advances}, 10\penalty0 (37):\penalty0 eadq1604, 2024.

\bibitem[Schatten et~al.(1969)Schatten, Wilcox, and Ness]{schatten1969model}
K.~H. Schatten, J.~M. Wilcox, and N.~F. Ness.
\newblock \emph{Solar Physics}, 6\penalty0 (3):\penalty0 442--455, 1969.

\bibitem[Schmahl et~al.(1982)Schmahl, Kundu, Strong, Bentley, Smith~Jr, and Krall]{schmahl1982active}
E.~Schmahl et al.
\newblock \emph{Solar Physics}, 80\penalty0 (2):\penalty0 233--249, 1982.

\bibitem[Seethapuram~Sridhar et~al.(2025)Seethapuram~Sridhar, Breen, Arumugam, Broderick, Cunningham, Green, Chrysostomou, and Ball]{seethapuram_sridhar_2025}
S.~Seethapuram~Sridhar et al.
\newblock A year in the life of the ska telescopes: overview and main outcomes, June 2025.
\newblock URL \url{https://doi.org/10.5281/zenodo.16950982}.

\bibitem[Segre and Zanza(2001)]{segre2001evolution}
S.~Segre and V.~Zanza.
\newblock \emph{The Astrophysical Journal}, 554\penalty0 (1):\penalty0 408, 2001.

\bibitem[Shibasaki et~al.(1994)Shibasaki, Enome, Nakajima, Nishio, Takano, Hanaoka, Torii, Sekiguchi, Kawashima, Bushimata, et~al.]{shibasaki1994purely}
K.~Shibasaki et al.
\newblock \emph{PASJ: Publications of the Astronomical Society of Japan (ISSN 0004-6264), vol. 46, no. 2, p. L17-L20}, 46:\penalty0 L17--L20, 1994.

\bibitem[Shimojo et~al.(2024)Shimojo, Bastian, Kameno, and Hales]{shimojo2024observing}
M.~Shimojo, T.~S. Bastian, S.~Kameno, and A.~S. Hales.
\newblock \emph{Solar Physics}, 299\penalty0 (2):\penalty0 20, 2024.

\bibitem[{Sim{\~o}es} and {Costa}(2010)]{simoes_2020}
P.~J.~A. {Sim{\~o}es} and J.~E.~R. {Costa}.
\newblock \emph{\solphys}, 266\penalty0 (1):\penalty0 109--121, Sept. 2010.
\newblock \doi{10.1007/s11207-010-9596-2}.

\bibitem[{Smerd} et~al.(1974){Smerd}, {Sheridan}, and {Stewart}]{Smerd1974}
S.~F. {Smerd}, K.~V. {Sheridan}, and R.~T. {Stewart}.
\newblock In G.~A. {Newkirk}, editor, \emph{Coronal Disturbances}, volume~57 of \emph{IAU Symposium}, page 389, January 1974.

\bibitem[{Smerd} et~al.(1975){Smerd}, {Sheridan}, and {Stewart}]{Smerd1975}
S.~F. {Smerd}, K.~V. {Sheridan}, and R.~T. {Stewart}.
\newblock \emph{\apjl}, 16\penalty0 (1):\penalty0 23--28, January 1975.

\bibitem[Stanislavsky et~al.(2015)Stanislavsky, Konovalenko, Koval, Dorovskyy, Zarka, and Rucker]{stanislavsky2015coronal}
A.~Stanislavsky et al.
\newblock \emph{Solar Physics}, 290\penalty0 (1):\penalty0 205--218, 2015.

\bibitem[Stupishin et~al.(2018)Stupishin, Kaltman, Bogod, and Yasnov]{stupishin2018modeling}
A.~Stupishin, T.~Kaltman, V.~Bogod, and L.~Yasnov.
\newblock \emph{Solar physics}, 293\penalty0 (1):\penalty0 13, 2018.

\bibitem[Tan et~al.(2012)Tan, Yan, Tan, Sych, and Gao]{tan2012microwave}
B.~Tan et al.
\newblock \emph{The Astrophysical Journal}, 744\penalty0 (2):\penalty0 166, 2012.

\bibitem[{Tingay} et~al.(2013){Tingay}, {Goeke}, {Bowman}, {Emrich}, {Ord}, {Mitchell}, {Morales}, {Booler}, {Crosse}, {Wayth}, {Lonsdale}, {Tremblay}, {Pallot}, {Colegate}, {Wicenec}, {Kudryavtseva}, {Arcus}, {Barnes}, {Bernardi}, {Briggs}, {Burns}, {Bunton}, {Cappallo}, {Corey}, {Deshpande}, {Desouza}, {Gaensler}, {Greenhill}, {Hall}, {Hazelton}, {Herne}, {Hewitt}, {Johnston-Hollitt}, {Kaplan}, {Kasper}, {Kincaid}, {Koenig}, {Kratzenberg}, {Lynch}, {Mckinley}, {Mcwhirter}, {Morgan}, {Oberoi}, {Pathikulangara}, {Prabu}, {Remillard}, {Rogers}, {Roshi}, {Salah}, {Sault}, {Udaya-Shankar}, {Schlagenhaufer}, {Srivani}, {Stevens}, {Subrahmanyan}, {Waterson}, {Webster}, {Whitney}, {Williams}, {Williams}, and {Wyithe}]{Tingay2013_MWA}
S.~J. {Tingay} et al.
\newblock \emph{\pasa}, 30:\penalty0 e007, Jan. 2013.
\newblock \doi{10.1017/pasa.2012.007}.

\bibitem[Tripathi et~al.(2023)Tripathi, Chakrabarty, Nandi, Prasad, Ramaprakash, Shaji, Sankarasubramanian, Thampi, and Yadav]{Tripathi2023AdityaL1}
D.~Tripathi et al.
\newblock \emph{Proceedings of the International Astronomical Union}, 18\penalty0 (S372):\penalty0 17--27, 2023.
\newblock \doi{10.1017/S1743921323001230}.

\bibitem[{Tun} and {Vourlidas}(2013)]{TunVourlidas2013}
S.~D. {Tun} and A.~{Vourlidas}.
\newblock \emph{\apj}, 766\penalty0 (2):\penalty0 130, April 2013.
\newblock \doi{10.1088/0004-637X/766/2/130}.

\bibitem[{Tzatzakis} et~al.(2008){Tzatzakis}, {Nindos}, and {Alissandrakis}]{tzatzakis_2008}
V.~{Tzatzakis}, A.~{Nindos}, and C.~E. {Alissandrakis}.
\newblock \emph{\solphys}, 253\penalty0 (1-2):\penalty0 79--94, Dec. 2008.
\newblock \doi{10.1007/s11207-008-9263-z}.

\bibitem[van~der Holst et~al.(2014)van~der Holst, Sokolov, Meng, Jin, Manchester~IV, Toth, and Gombosi]{van2014alfven}
B.~van~der Holst et al.
\newblock \emph{The Astrophysical Journal}, 782\penalty0 (2):\penalty0 81, 2014.

\bibitem[{van Haarlem} et~al.(2013){van Haarlem}, {Wise}, {Gunst}, {Heald}, {McKean}, {Hessels}, {de Bruyn}, {Nijboer}, {Swinbank}, {Fallows}, {Brentjens}, {Nelles}, {Beck}, {Falcke}, {Fender}, {H{\"o}randel}, {Koopmans}, {Mann}, {Miley}, {R{\"o}ttgering}, {Stappers}, {Wijers}, {Zaroubi}, {van den Akker}, {Alexov}, {Anderson}, {Anderson}, {van Ardenne}, {Arts}, {Asgekar}, {Avruch}, {Batejat}, {B{\"a}hren}, {Bell}, {Bell}, {van Bemmel}, {Bennema}, {Bentum}, {Bernardi}, {Best}, {B{\^\i}rzan}, {Bonafede}, {Boonstra}, {Braun}, {Bregman}, {Breitling}, {van de Brink}, {Broderick}, {Broekema}, {Brouw}, {Br{\"u}ggen}, {Butcher}, {van Cappellen}, {Ciardi}, {Coenen}, {Conway}, {Coolen}, {Corstanje}, {Damstra}, {Davies}, {Deller}, {Dettmar}, {van Diepen}, {Dijkstra}, {Donker}, {Doorduin}, {Dromer}, {Drost}, {van Duin}, {Eisl{\"o}ffel}, {van Enst}, {Ferrari}, {Frieswijk}, {Gankema}, {Garrett}, {de Gasperin}, {Gerbers}, {de Geus}, {Grie{\ss}meier}, {Grit}, {Gruppen}, {Hamaker}, {Hassall}, {Hoeft}, {Holties}, {Horneffer},
  {van der Horst}, {van Houwelingen}, {Huijgen}, {Iacobelli}, {Intema}, {Jackson}, {Jelic}, {de Jong}, {Juette}, {Kant}, {Karastergiou}, {Koers}, {Kollen}, {Kondratiev}, {Kooistra}, {Koopman}, {Koster}, {Kuniyoshi}, {Kramer}, {Kuper}, {Lambropoulos}, {Law}, {van Leeuwen}, {Lemaitre}, {Loose}, {Maat}, {Macario}, {Markoff}, {Masters}, {McFadden}, {McKay-Bukowski}, {Meijering}, {Meulman}, {Mevius}, {Middelberg}, {Millenaar}, {Miller-Jones}, {Mohan}, {Mol}, {Morawietz}, {Morganti}, {Mulcahy}, {Mulder}, {Munk}, {Nieuwenhuis}, {van Nieuwpoort}, {Noordam}, {Norden}, {Noutsos}, {Offringa}, {Olofsson}, {Omar}, {Orr{\'u}}, {Overeem}, {Paas}, {Pandey-Pommier}, {Pandey}, {Pizzo}, {Polatidis}, {Rafferty}, {Rawlings}, {Reich}, {de Reijer}, {Reitsma}, {Renting}, {Riemers}, {Rol}, {Romein}, {Roosjen}, {Ruiter}, {Scaife}, {van der Schaaf}, {Scheers}, {Schellart}, {Schoenmakers}, {Schoonderbeek}, {Serylak}, {Shulevski}, {Sluman}, {Smirnov}, {Sobey}, {Spreeuw}, {Steinmetz}, {Sterks}, {Stiepel}, {Stuurwold}, {Tagger}, {Tang},
  {Tasse}, {Thomas}, {Thoudam}, {Toribio}, {van der Tol}, {Usov}, {van Veelen}, {van der Veen}, {ter Veen}, {Verbiest}, {Vermeulen}, {Vermaas}, {Vocks}, {Vogt}, {de Vos}, {van der Wal}, {van Weeren}, {Weggemans}, {Weltevrede}, {White}, {Wijnholds}, {Wilhelmsson}, {Wucknitz}, {Yatawatta}, {Zarka}, {Zensus}, and {van Zwieten}]{lofar2013}
M.~P. {van Haarlem} et al.
\newblock \emph{\aap}, 556:\penalty0 A2, Aug. 2013.
\newblock \doi{10.1051/0004-6361/201220873}.

\bibitem[{Vasanth} et~al.(2016){Vasanth}, {Chen}, {Feng}, {Ma}, {Du}, {Song}, {Kong}, and {Wang}]{vasanth2016}
V.~{Vasanth} et al.
\newblock \emph{\apjl}, 830\penalty0 (1):\penalty0 L2, Oct. 2016.
\newblock \doi{10.3847/2041-8205/830/1/L2}.

\bibitem[{Vr{\v{s}}nak} et~al.(2001){Vr{\v{s}}nak}, {Aurass}, {Magdaleni{\'c}}, and {Gopalswamy}]{vrsnak2001}
B.~{Vr{\v{s}}nak}, H.~{Aurass}, J.~{Magdaleni{\'c}}, and N.~{Gopalswamy}.
\newblock \emph{\aap}, 377:\penalty0 321--329, October 2001.
\newblock \doi{10.1051/0004-6361:20011067}.

\bibitem[{Vr{\v{s}}nak} et~al.(2002){Vr{\v{s}}nak}, {Magdaleni{\'c}}, {Aurass}, and {Mann}]{Vrsnak2002}
B.~{Vr{\v{s}}nak}, J.~{Magdaleni{\'c}}, H.~{Aurass}, and G.~{Mann}.
\newblock \emph{\aap}, 396:\penalty0 673--682, Dec. 2002.
\newblock \doi{10.1051/0004-6361:20021413}.

\bibitem[{Wang} et~al.(2003){Wang}, {Duan}, {Xie}, and {Yan}]{Wang2003}
M.~{Wang}, C.~C. {Duan}, R.~X. {Xie}, and Y.~H. {Yan}.
\newblock \emph{\solphys}, 212\penalty0 (2):\penalty0 401--406, Feb. 2003.
\newblock \doi{10.1023/A:1022939203483}.

\bibitem[White and Kundu(1997)]{white1997radio}
S.~White and M.~Kundu.
\newblock \emph{Solar Physics}, 174\penalty0 (1):\penalty0 31--52, 1997.

\bibitem[White(2004)]{white2004coronal}
S.~M. White.
\newblock \emph{Solar and Space Weather Radiophysics: Current Status and Future Developments}, pages 89--113, 2004.

\bibitem[Wiegelmann(2008)]{wiegelmann2008nonlinear}
T.~Wiegelmann.
\newblock \emph{Journal of Geophysical Research: Space Physics}, 113\penalty0 (A3), 2008.

\bibitem[Wiegelmann and Sakurai(2021)]{wiegelmann2021solar}
T.~Wiegelmann and T.~Sakurai.
\newblock \emph{Living Reviews in Solar Physics}, 18\penalty0 (1):\penalty0 1, 2021.

\bibitem[Wiegelmann et~al.(2017)Wiegelmann, Petrie, and Riley]{wiegelmann2017coronal}
T.~Wiegelmann, G.~J. Petrie, and P.~Riley.
\newblock \emph{Space Science Reviews}, 210\penalty0 (1):\penalty0 249--274, 2017.

\bibitem[Wild and Smerd(1972)]{wild1972radio}
J.~Wild and S.~Smerd.
\newblock \emph{Annual Review of Astronomy and Astrophysics, vol. 10, p. 159}, 10:\penalty0 159, 1972.

\bibitem[{Wild}(1967)]{culgoora_1967}
J.~P. {Wild}.
\newblock \emph{\pasa}, 1\penalty0 (2):\penalty0 38--39, Nov. 1967.
\newblock \doi{10.1017/S1323358000010407}.

\bibitem[Yan et~al.(2021)Yan, Chen, Wang, Liu, Geng, Chen, Tan, Chen, Su, and Tan]{yan2021mingantu}
Y.~Yan et al.
\newblock \emph{frontiers in Astronomy and Space Sciences}, 8:\penalty0 584043, 2021.

\bibitem[{Zhelezniakov} and {Zlotnik}(1980)]{zhelezniakov_and_zlotnik}
V.~V. {Zhelezniakov} and E.~I. {Zlotnik}.
\newblock In M.~R. {Kundu} and T.~E. {Gergely}, editors, \emph{Radio Physics of the Sun}, volume~86 of \emph{IAU Symposium}, pages 87--99, Jan. 1980.
\newblock \doi{10.1017/S0074180900036676}.

\bibitem[Zheleznyakov(1962)]{zheleznyakov1962origin}
V.~Zheleznyakov.
\newblock \emph{Soviet Astronomy, Vol. 6, p. 3}, 6:\penalty0 3, 1962.

\bibitem[Zheleznyakov(1970)]{zheleznyakov1970radio}
V.~Zheleznyakov.
\newblock \emph{International Series of Monographs in Natural Philosophy}, 1970.

\bibitem[Zheleznyakov and Zlotnik(1964)]{zheleznyakov1964polarization}
V.~Zheleznyakov and E.~Y. Zlotnik.
\newblock \emph{Soviet Astronomy, Vol. 7, p. 485}, 7:\penalty0 485, 1964.

\bibitem[Zheleznyakov and Zlotnik(1977)]{zheleznyakov1977propagation}
V.~Zheleznyakov and E.~Y. Zlotnik.
\newblock \emph{Radiophysics and Quantum Electronics}, 20\penalty0 (9):\penalty0 997--1009, 1977.

\bibitem[Zhukov et~al.(2025)Zhukov, Thizy, Galano, Bourgoignie, Dolla, Jean, Nicula, Shestov, Galy, Rougeot, Versluys, Zender, Lamy, Fineschi, Gunár, Inhester, Mierla, Rudawy, Steslicki, Zangrilli, and et~al.]{Zhukov2025ASPIICS}
A.~N. Zhukov et al.
\newblock \emph{arXiv e-prints}, \penalty0 (arXiv:2509.00253), 2025.
\newblock \doi{10.48550/arXiv.2509.00253}.

\bibitem[{Zimovets} et~al.(2012){Zimovets}, {Vilmer}, {Chian}, {Sharykin}, and {Struminsky}]{Zimovets2012}
I.~{Zimovets} et al.
\newblock \emph{\aap}, 547:\penalty0 A6, Nov. 2012.
\newblock \doi{10.1051/0004-6361/201219454}.

\bibitem[{Zimovets} and {Sadykov}(2015)]{Zimovets2015}
I.~V. {Zimovets} and V.~M. {Sadykov}.
\newblock \emph{Advances in Space Research}, 56\penalty0 (12):\penalty0 2811--2832, Dec. 2015.
\newblock \doi{10.1016/j.asr.2015.01.041}.

\bibitem[{Zlotnik}(1981)]{Zlotnik1981}
E.~I. {Zlotnik}.
\newblock \emph{\aap}, 101\penalty0 (2):\penalty0 250--258, Aug. 1981.

\bibitem[{Zlotnik}(1968)]{zlotnik_ff}
E.~Y. {Zlotnik}.
\newblock \emph{\sovast}, 12:\penalty0 245, Oct. 1968.

\bibitem[{Zucca} et~al.(2018){Zucca}, {Morosan}, {Rouillard}, {Fallows}, {Gallagher}, {Magdalenic}, {Klein}, {Mann}, {Vocks}, {Carley}, {Bisi}, {Kontar}, {Rothkaehl}, {Dabrowski}, {Krankowski}, {Anderson}, {Asgekar}, {Bell}, {Bentum}, {Best}, {Blaauw}, {Breitling}, {Broderick}, {Brouw}, {Br{\"u}ggen}, {Butcher}, {Ciardi}, {de Geus}, {Deller}, {Duscha}, {Eisl{\"o}ffel}, {Garrett}, {Grie{\ss}meier}, {Gunst}, {Heald}, {Hoeft}, {H{\"o}randel}, {Iacobelli}, {Juette}, {Karastergiou}, {van Leeuwen}, {McKay-Bukowski}, {Mulder}, {Munk}, {Nelles}, {Orru}, {Paas}, {Pandey}, {Pekal}, {Pizzo}, {Polatidis}, {Reich}, {Rowlinson}, {Schwarz}, {Shulevski}, {Sluman}, {Smirnov}, {Sobey}, {Soida}, {Thoudam}, {Toribio}, {Vermeulen}, {van Weeren}, {Wucknitz}, and {Zarka}]{Zucca2018}
P.~{Zucca} et al.
\newblock \emph{\aap}, 615:\penalty0 A89, July 2018.
\newblock \doi{10.1051/0004-6361/201732308}.

\bibitem[{Zucca} et~al.(2025){Zucca}, {Zhang}, {Kozarev}, {Nedal}, {Dey}, {Mancini}, {Kumari}, {Morosan}, {Dabrowski}, {Gallagher}, {Krankowski}, and {Vocks}]{Zucca2025}
P.~{Zucca} et al.
\newblock \emph{\aap}, 703:\penalty0 A271, Nov. 2025.
\newblock \doi{10.1051/0004-6361/202554348}.

\end{thebibliography}

\end{document}